\documentclass[useAMS,usenatbib]{mn2e}

\usepackage{graphicx}
\usepackage{remreset} 
\usepackage{multicol}
\usepackage{amsmath}

\def\apj{ApJ}
\def\mnras{MNRAS}
\def\aap{A\&A}
\def\apjs{ApJS}
\def\apjl{ApJL}
\def\aj{AJ}

\title[Testing \textit{WMAP} data via \textit{Planck}]{Testing \textit{WMAP} data via \textit{Planck} radio and SZ catalogues}
\author[J.R. Whitbourn, T. Shanks, U. Sawangwit]{J.R. Whitbourn$^{1}$
\thanks{Corresponding author:(JRW) joseph.whitbourn@durham.ac.uk}, T. Shanks$^{1}$ and U. Sawangwit$^{1,2}$\\
$^{1}$Department of Physics, Durham University, South Road DH1 3LE\\
$^{2}$ National Astronomical Research Institute of Thailand, Thailand\\}

\begin{document}

\date{Accepted \textbf{}; Received \textbf{}; in original form \textbf{}}
\pagerange{\pageref{firstpage}--\pageref{lastpage}} \pubyear{2011}

\maketitle

\label{firstpage}


\begin{abstract}

The prime evidence underpinning the standard $\Lambda$CDM cosmological
model is the CMB power spectrum as observed by \textit{WMAP} and other
microwave experiments. But \citet{utaneradio} have recently shown that
the \textit{WMAP} CMB power spectrum is highly sensitive to the beam
profile of the \textit{WMAP} telescope. Here, we use the source
catalogue from the \textit{Planck} Early Data Release to test further
the \textit{WMAP} beam profiles. We confirm that stacked beam profiles
at Q, V and particularly at W appear wider than expected when compared
to the Jupiter beam, normalised either directly to the radio source
profiles or using \textit{Planck} fluxes. The same result is also found
based on \textit{WMAP}-CMBfree source catalogues and NVSS sources. The accuracy
of our beam profile measurements is supported by analysis of CMB sky
simulations. However the beam profiles from \textit{WMAP}7 at the W band are
narrower than previously found in \textit{WMAP}5 data and the rejection of the
\textit{WMAP} beam is now only at the $\approx3\sigma$ level. We also find that
the \textit{WMAP} source fluxes demonstrate possible non-linearity with
\textit{Planck} fluxes. But including ground-based and \textit{Planck} data
for the bright \citet{weiland2011} sources may suggest that the discrepancy is 
a linear offset rather than a non-linearity.  Additionally, we find that the
stacked Sunyaev-Zel'dovich (SZ) decrements of $\approx151$ galaxy clusters
observed by \textit{Planck} are in agreement with the \textit{WMAP} data. We
find that there is no evidence for a \textit{WMAP} SZ deficit as has previously
been reported. In the particular case of Coma we find evidence for the presence
of an $\mathcal{O}(0.1mK)$ downwards CMB fluctuation. We conclude that beam
profile systematics can have significant effects on both the amplitude and
position of the acoustic peaks, with potentially important implications for
cosmology parameter fitting.

\end{abstract}




\section{Introduction}
\label{sec:intro}

Cosmic Microwave Background (CMB) experiments such as the Wilkinson Microwave
Anisotropy Probe (\textit{WMAP}) have made significant progress in the study
of the primordial temperature fluctuations. Their best fitting power
spectra strongly support a spatially flat, $\Lambda CDM$, universe. This
model requires relatively few  parameters, yet apparently manages a
compelling concordance between a variety of other cosmological data;
SNIa, Large Scale Structure and Big Bang Nucleosynthesis. 
Although the statistical errors on these power spectra are small, this
precision does not necessarily imply accuracy and there remains the
potential for systematic errors to alter these conclusions.

Indeed, several anomalies between $\Lambda CDM$ and the \textit{WMAP} data have
been discussed. Typically these have involved the large-scale temperature
multipoles eg: \citep{bennett2011,liuli}. However, other anomalies in the CMB at
smaller scales have also been detected, connected in particular with
radio sources \citep{utaneradio, tommoriond} and SZ decrements from galaxy clusters
\citep{myers2004,bielby2007}

Radio sources are sometimes  regarded as a contaminant in  CMB
temperature maps. However, radio point sources prove particularly
interesting because they provide a complementary check of the beam
measured by the \textit{WMAP} team from observations of Jupiter
\citep{page2003,hill2009}. Jupiter has a flux of $\approx$ 1200Jy which
is $\approx3$ orders of magnitude higher than radio source fluxes or CMB
fluctuations. This high flux has advantages in terms of defining the
wings of the beam profile but has the disadvantage that the calibrating
source is much brighter than typical CMB fluctuations. Furthermore,
Jupiter only checks the beam on the ecliptic whereas radio sources are
spread over the sky. \citet{utaneradio,tommoriond} made a stacked
analysis of radio point sources and found evidence for a wider beam than
\textit{WMAP} measured using Jupiter.  A tentative detection of a
non-linear relation between \textit{WMAP} fluxes and ground based radio
telescope fluxes was also found. A thorough analysis of possible
systematics did not find an explanation and we return to these issues
later in this paper. The beam profile of a CMB telescope  like \textit{WMAP} is critical
because it smoothes the temperature anisotropies and therefore needs to
be known accurately to produce the final power spectrum from 
temperature maps \citep{page2003,hill2009}. 

Various authors have noted small-scale anomalies with respect to the SZ
decrements measured by \textit{WMAP}. SZ decrements are created when CMB photons
inverse Compton scatter off hot electrons in galaxy clusters.
\citet{myers2004} first stacked \textit{WMAP} data at the positions of galaxy
clusters and suggested that the profiles were more extended than
expected. \citet{lieu2006} and \citet{bielby2007} then found that
the SZ decrements from \textit{WMAP} were reduced compared to X-ray
predictions, possibly due to the \textit{WMAP} beam being wider than
expected. \citet{bielby2007} also found that the \textit{WMAP} decrements were 
significantly lower than the ground-based SZ measurements by
\citet{bonamente2006} in 38 X-ray luminous clusters.

In their ESZ sample, the \textit{Planck} team find excellent agreement
with the self-similar X-ray estimates of the SZ decrement
\citep{planckxraystats2011,planckszstudy2011}. This is corroborated by the ground based
South Pole Telescope Collaboration with their blind SZ selected  cluster
sample \citep{southpole2009}. This compounds the question of why
\textit{WMAP} SZ analyses from \citet{lieu2006} and \citet{bielby2007}
failed to find such an agreement.

In this paper we use the recent \textit{Planck} Early Data Release and other radio
source data to re-investigate both the \textit{WMAP} radio source beam
profile and SZ anomalies. The \textit{Planck} Early Release Compact Source
catalogue (ERCSC) is of particular interest and provides the basic
parameters of radio sources and SZ clusters from the \textit{Planck} CMB maps.
Although, the corresponding temperature maps from which these were
estimated have not been released, both radio source fluxes and SZ
profile parameters are available as measured by \textit{Planck}. We can therefore
use these to compare \textit{WMAP} and \textit{Planck} radio source fluxes
directly and also to make  \textit{WMAP} stacks centred now on the new
radio source and SZ cluster lists from \textit{Planck}. From these stacks, the
\textit{WMAP} beam profile can be inferred and the SZ results from
\textit{WMAP} and \textit{Planck} compared. Given the higher angular resolution,
lower noise and different calibration strategy for \textit{Planck}, this
comparison will allow new insight into the robustness of
the \textit{WMAP} CMB analysis.

\section{Data}
\label{sec:data}

\subsection{\textit{Planck} Early Data Release}

The \textit{Planck} team have recently made their first release of data
collected by the \textit{Planck} satellite between 13 August 2009 and 6 June 2010
(amounting to $\approx 1.5$ full sky surveys). This early data release
is concerned solely with the foreground contamination in the CMB maps.
The two sets of catalogues relevant to this paper form the Early Release
Compact Source catalogue (ERCSC). These are the Radio Source catalogues and the
SZ  catalogue.

\subsubsection{\textit{Planck} Radio Sources}

The Early Release Compact Source  Catalogue (ERCSC) lists all the high
reliability radio sources with accurate flux determinations. The ERCSC
has been quality controlled so that $\ge 90\%$ of the reported sources
are reliable, $>5\sigma$, detections and that the fluxes are determined
within $\le$30\% accuracy. The catalogues are band specific and for the
bands of interest ($\nu \leq 100GHz$) are created using the
`PowellSnakes' method, a Bayesian multi-frequency algorithm for
detecting discrete objects in a random background. Flux estimates were
obtained by use of aperture photometry within a circle of the beam's
FWHM. For the case of unresolved and potentially faint point sources,
the \textit{Planck} team recommend the use of the parameter FLUX and its
corresponding error, FLUX\_ERR \citep{planckERCSCexplain}. 

We reject any extended objects from the catalogue to maintain an
unresolved sample with which to test the \textit{WMAP} data. To
do this we have used the \textit{Planck} quality tag `EXTENDED'. This
is defined by comparing the source areal profile with the 2-D
\textit{Planck} beam. An additional quality flag
`CMBSUBTRACT' has also been provided, which reflects on the quality of
the source detection in a map with the best estimate of the CMB removed.
We minimise CMB contamination by using only CMBSUBTRACT=0 sources.

When measuring the beam profile in Section \ref{subsec:planckradioresults} we
further cut the catalogue to ensure the best quality sample. \citet{utaneradio}
did suggest that their faintest \textit{WMAP} source samples were probably
affected by \citet{eddington1913} bias. To ensure the robustness of our results
against Eddington bias, we have used a $S\geq1.1$Jy flux cut, the same limit as
previously used by \citet{utaneradio}. We have additionally rejected sources
within $4^\circ$ of the LMC, sources at low galactic latitude, $|b|<5^\circ$ and
any sources flagged by \textit{Planck} as having high astrometric error.
Finally, we tightened the \textit{Planck} `EXTENDED' flag to remove any sources
intrinsically wider than the \textit{WMAP} beam. The \textit{Planck} `EXTENDED'
flag excludes sources with $(GAU_{FWHM\_MAJ}\times GAU_{FWHM\_MIN})^{1/2} >
1.5\times(BEAM_{FWHM\_MAJ}\times BEAM_{FWHM\_MIN})^{1/2}$. We now ensure 
that the \textit{Planck} sources are unresolved in the \textit{WMAP} maps by
imposing cuts in both the major and minor axis so that both the fitted
Gaussian profiles (GAU) and the local PSF (BEAM) FWHM estimates are less than
the FWHM of the \textit{WMAP} beam in the band being studied\footnote{We relax
this cut for the Q-band, here we only impose cuts on local PSF (BEAM) FWHM
estimates to ensure we get a reasonable number of sources.}.

Band and colour corrections for the \textit{WMAP} and \textit{Planck} fluxes have been
ignored. This factor is in any case small due to the typically flat spectral
indices considered \citep{planckERCSCexplain, wright2009}. The full
details of the catalogue construction and composition are described by
\citet{planckearlySZ} and briefly overviewed in Table \ref{tb:bands-planckpara}.

\subsubsection{\textit{Planck} SZ Catalogue}

The Early SZ (ESZ) catalogue lists all the robust and extensively
verified SZ detections in the first data release. As described  by
\citet{melin2006}, the \textit{Planck} team extract the integrated SZ signal,
the \textit{Y} parameter, using a Multifrequency Matched filter (MMF3) method
\citep{planckearlySZ}. The algorithm is run blindly on all-sky maps,
assuming the characteristic SZ spectral signature and self-similar
cluster profile.

In the Early Release of the \textit{Planck} SZ catalogue, only data from
the 100GHz frequency channel or higher has been used to study the SZ
effect. This is to avoid the detrimental effect on S/N from beam
dilution caused by the larger beam sizes of the lower frequency
channels. At the higher frequencies, the \textit{Planck} beam FWHM is
typically $\approx4.'5$. The full details of the catalogue construction
and composition are described by the \citet{planckearlySZ}.

The catalogue provides estimates of the SZ flux, extent, redshift and
position. It consists of 189 clusters, all detected at high S/N ($\ge6$)
with 95\% reliability. Whilst the sample is primarily composed of known
clusters (169/189), it provides a wealth of new information as it gives
the first SZ measurements for $\approx 80\%$ of the clusters. In this
paper we only make use of clusters which have been pre-detected in the X-ray 
and have redshifts. We therefore, after masking, consider 151 clusters, including
Coma. For this sample the redshift range spans $z \in [0.0126,0.546]$
with a mean redshift of $\overline{z} = 0.18$.

\subsection{\textit{WMAP} Data}
\label{sec:wmap_data}

We will be using the 7-year \textit{WMAP} temperature maps obtained from
the LAMBDA CMB resource. We work with the $N_{side} = 512$
\texttt{HEALPIX} maps resulting in a pixel scale of $7'$. We use the
foreground unsubtracted temperature band maps for  Q,V and W. Our
default \textit{WMAP} datasets are the co-added maps in Q (=Q1+Q2), V (=V1+V2)
and W (=W1+W2+W3+W4). However, particularly in the W band, the increased
S/N for radio source profiles obtained by using all the DA's can be
regarded as a trade-off with the precision of just using the narrowest
W1 (and W4) beams as previously used by \citet{utaneradio}. In using the
co-added data, the Jupiter beams have to be combined before comparison
with the data. We estimate the Jupiter beam in each band by averaging
the 7-year beam profiles from the various detector assemblies, assuming
the appropriate correction for pixelisation \citep{hinshaw03}. When
working with radio point sources we  use the point source catalogue mask
(\texttt{wmap\_point\_source\_catalog\_mask}). To avoid Galactic  contamination
for the SZ analyses we have instead used the extended temperature mask
(\texttt{wmap\_ext\_temperature\_analysis\_mask}) which admits 71\% of the sky.

We have used the 7-year \textit{WMAP} 5-band point source catalogue
\citep{gold2011}. These sources are detected at least the 5$\sigma$ level in one
\textit{WMAP} band. For a flux density to be stated, the detection must be above
the 2$\sigma$ level in that band. Following \citet{utaneradio} we ensure
that the sources are genuinely point sources by matching to the 5GHz
($\approx4'.6$ resolution) catalogues from the Greenbank Northern sky
Survey (GB6, \citealt{gregory1996}), or Parkes-MIT-NRAO (PMN,
\citealt{griffith1993}), surveys. The \textit{WMAP} team also provide a 7
yr CMB-free catalogue as described by \citet{gold2011}. This catalogue
has been created with the objective of detecting point sources free of
boosting by CMB fluctuations. We proceed with the raw 5-band catalogue with 471
sources and a CMB-free catalogue with 417
sources.


\begin{table}
\begin{center}
\begin{tabular}{cccc}

\hline\hline
 Freq & FWHM & Flux Limit \\

[GHz] & ($\arcmin$) & [Jy] \\
\hline

100 & 9.94 & 0.344 \\

70 & 13.01 & 0.481 \\

44 & 27.00 & 0.781 \\

\hline\hline
\end{tabular}
\caption{Summary of the \textit{Planck} bandpass parameters and the flux range of
the sample we use from the ERCSC, \citep{planckERCSCexplain}.
}
\end{center}
\label{tb:bands-planckpara}
\end{table}


\begin{table}
 \begin{tabular}{cccccc}

\hline\hline
Band & Freq & FWHM & $\Omega$ & $\Gamma^{ff}$ & $g(\nu)$ \\

 & [GHz] & ($\arcmin$) & (sr) & $ [\mu K Jy^{-1}]$ & $ $ \\
\hline
W & 94 & 12.6 & 2.097$\cdot10^{-5}$ & 179.3 & 1.245 \\
V & 61 & 19.8 & 4.202$\cdot10^{-5}$ & 208.6 & 1.099 \\
Q & 41 & 29.4 & 8.978$\cdot10^{-5}$ & 216.6 & 1.044 \\
%
\hline\hline
\end{tabular}
\caption{Summary of the \textit{WMAP} bandpass parameters  taken from \citet{hill2009} and
\citet{jarosik2011}. See text for definitions.
}
\label{tb:bands-wmappara}
\end{table}


\section{\textit{Planck} radio source fluxes and SZ cluster decrements}
\label{sec:model-radio}

\subsection{Conversion of Radio Flux to Temperature Profiles}
\label{subsec:model-radiofluxes}

The \textit{Planck} ERCSC provides us with the source flux density,  error
and a few parameters on the source characteristics and detection. To
enable us to translate the \textit{Planck} fluxes into \textit{WMAP} observables we need
to convert the source flux density, $S_{tot}$, into an observed peak Rayleigh-Jeans antenna
temperature using the conversion factor $\Gamma^{ff}(\nu)$ \citep{page2003preflight},

\begin{equation}
 \Delta T_{RJ}(0) = S_{tot} \Gamma^{ff}(\nu),
\label{eq:gammatemprel}
\end{equation}

\noindent where

\begin{equation}
 \Gamma^{ff}(\nu) = \frac{c^{2}}{2k_{b}\nu^{2}_{e}} \frac{1}{\Omega_{beam}(\nu)}.
\label{eq:gammadef}
\end{equation}

\noindent Here $\nu_{e}$ is the effective frequency of the bandpass and the \textit{ff} superscript denotes the
fact that the majority of the \textit{WMAP} sources have a spectral index $\alpha
\approx -0.1$, approximately that of free-free emission.

The \textit{WMAP} temperature maps are given in terms of the thermodynamic
temperature. At the \textit{WMAP} frequencies and CMB temperature, the Rayleigh-Jeans
temperature is appreciably different from this. We therefore correct between the two
temperature differences, using eq({\ref{eq:anttothermo}}), where
$x'=h\nu/k_{b}T_{cmb}$ and $T_{cmb}=2.725$K is the monopole temperature
of the CMB \citep{jarosik2003}.

\begin{align}
 \Delta T_{t} &= \frac{(e^{x'}-1)^{2}}{x'^{2}e^{x'}} \Delta T_{RJ} \label{eq:anttothermo}, \\
 &= g(\nu) \Delta T_{RJ} \nonumber.
\end{align}

\noindent The observed \textit{WMAP} temperature profiles therefore take the form,

\begin{align}
 \Delta T(\theta) &= \Delta T(0) b^{s}(\theta) \label{eq:radiotempprofile}, \\
 &= g(\nu)\Gamma^{ff} S_{tot} b^{s}(\theta) \nonumber.
\end{align}

\noindent We see the beam dependence of the observed profile is twofold.
The shape is dependent on the symmetrized beam profile $b^{s}(\theta)$
(normalised to unity at $\theta=0^{\circ}$), while the scale is
normalised by the beam solid angle associated with $\Gamma^{ff}$. A
summary of the assumed  values of $g(\nu)$ and $\Gamma^{ff}$ is
provided in Table {\ref{tb:bands-wmappara}}.


\subsection{\textit{Planck} SZ Decrements}
\label{sec:model-sz}

\textit{Planck} presents its observed decrements using an SZ model fit parameterised 
by the total SZ signal within the cluster extent. Here we briefly describe this
model so that the \textit{Planck} results can be compared to the stacked \textit{WMAP} temperature decrements. 

Clusters are significant reservoirs of gas which will result in a SZ
distortion to the CMB described by the Compton \textit{y} parameter,

\begin{equation}
 \Delta T(\theta) = T_{cmb} j(x') y(\theta).
\label{eq:szyandtemp}
\end{equation}

\noindent Here, $j(x')$ is the spectral function, where $x'=h\nu/k_{b}T_{cmb}$ 
(\citealt{sunyaev1980}), 

\begin{equation}
 j(x') = \frac{x'(e^{x'}+1)}{e^{x'}-1}-4.
\label{eq:spectralwritten}
\end{equation}

The integrated \textit{Y} parameter is the total
SZ signal, which is simply the integration of the Compton y parameter
on the sky,

\begin{equation}
 Y = \int y d\Omega.
\label{eq:totalY}
\end{equation}

\noindent Alternatively, if we integrate over the cluster volume,

\begin{equation}
 Y = \frac{\sigma_{t}}{m_{e}c^{2}} \int{P dV}.
\label{eq:totalYthreeD}
\end{equation}

\noindent However, we are observing  a 2-D projection of the
cluster\footnote{The cluster is assumed to be spherical.} on the sky.
The angle $\theta$ we observe on the sky, corresponds in 3-D to a
cylindrical bore through the cluster of radius $R = \theta D_a(z)$. where
$D_a$ is the angular diameter distance. The observed integrated
\textit{Y} parameter therefore takes the form \citep{arnaud2010},

\begin{align}
 Y_{cyl}(R) &= \frac{\sigma_{t}}{m_{e}c^{2}} \int^{R}_{0} 2\pi r dr \int^{R_{tot}}_{r} \frac{2P(r')r'dr'}{(r'^{2}-r^{2})^{1/2}}, \label{eq:rawcylprofile} \\
 &= Y_{sph}(R_{tot}) - \frac{\sigma_{t}}{m_{e}c^{2}} \int^{R_{tot}}_{R} 4 \pi P(r) (r^{2} - R^{2})^{1/2} r dr \notag.
\end{align}

To predict the SZ effect implied by eq({\ref{eq:rawcylprofile}}) we have
to make a choice of the pressure profile, $P(r)$. Historically it has
been common to fit the SZ profile with an isothermal $\beta$ model,
\citep{cavaliere1976}. However, X-ray observations have shown
that the assumption of an isothermal gas breaks down at the cluster
outskirts, \citep{pratt2007, piffaretti2005}. To account for
this additional complexity, \citet{nagai2007} therefore proposed using a
Generalised NFW (GNFW) profile for the pressure instead. The profile is
scale invariant in that it is independent of absolute distances and is
instead a function of the dimensionless scale $x=R/R_{500}$. The profile
takes the form,

\begin{equation}
 \mathcal{P}(x) = \frac {P_{0}}{(c_{500}x)^{\gamma}[1+(c_{500}x)^{\alpha}]^{(\beta-\gamma)/\alpha}},
\label{eq:nfwprofile}
\end{equation}

\noindent where $\mathcal{P}(x)=P(r)/P_{500}$ and $P_{500}$ is the 
characteristic pressure defined by \citet{arnaud2010}.

Here we have a five parameter fit to the pressure profile,
$[P_{0},c_{500},\gamma,\alpha,\beta]$. This allows independent
specification of the pressure in the cluster core ($\gamma$), main-body
($\alpha$) and outskirts ($\beta$). In Table {\ref{tb:nfw-planckpara}}
we outline the parameters used by \textit{Planck}, as taken from
\citet{arnaud2010}. The characteristic parameters of the cluster are
$M_{500}$, $P_{500}$, $R_{500}$ (see Appendix \ref{append:selfsim}) where the
500 denotes the fact they are evaluated within the region where the mean mass
density is 500 times greater that the critical density $\rho_{crit}(z)$. The
\textit{Planck} team extract the integrated \textit{Y} parameter using the
Multifrequency Matched Filter (MMF3) method \citep{planckearlySZ} based
on the above self-similar model. The integration is done to the angular cluster
extent corresponding to $5R_{500}$, which \textit{Planck} also report
($\theta_{5R500}$). Their errors on the integrated SZ signal, \textit{Y},
combine their estimated measurement error with Monte-Carlo estimates of the
error due to uncertainities within the self-similar model \citep{melin2006}.

\subsection{SZ Temperature Profile Reconstruction}
\label{subsec:model_sz_recon}

We now proceed to invert the \textit{Planck} data to provide us with
expected \textit{WMAP} temperature profiles. (See Appendix \ref{append:selfsim}
for the details of this derivation). From the \textit{Planck} values for 
$Y(5R_{500})$ and $\theta_{5R500}$, and using  $J(x)$ and $I(x)$,
the cylindrical and spherical SZ templates (see eq.(\ref{eq:jscaling})) 
we first obtain $Y_{cyl}(R)$ via eq.(\ref{eq:thereallyusefulY2}),

\begin{equation}
 Y_{cyl}(R) = Y_{cyl}(5R_{500}) \bigg( 1 - \frac{J(x)}{I(5)} \bigg).
\label{eq:thereallyusefulY}
\end{equation}


\noindent From this integrated $Y_{cyl}(R=\theta\cdot
D_A(z))$, we want to derive the angular dependence of the Compton \textit{y}
parameter, where $y(\theta) = \frac{d}{d\Omega}Y_{cyl}(\theta)$, and so 

\begin{equation}
 y(\theta) = -\frac{Y_{cyl}(5R_{500})}{I(5)} \frac{d}{d\Omega}\big(J(x)\big).
\label{eq:ythetaselfsim}
\end{equation}

\noindent The self-similar model therefore predicts an SZ temperature decrement,

\begin{equation}
 \Delta T_{SZ}(\theta) = - T_{cmb} j(x') \frac{Y_{cyl}(5R_{500})}{I(5)} \frac{d}{d\Omega}\big(J(x)\big),
\label{eq:selfsimtemppredfinal}
\end{equation}

\noindent where $Y_{cyl}(5R_{500})$ is the integrated $Y$ given in the ESZ.

\subsection{Convolution with the \textit{WMAP} beam profile}
\label{subsec:model_sz_convol}

\noindent The cluster profile is not directly observed by \textit{WMAP} and is
instead smoothed by the instrument response. The predicted \textit{WMAP} SZ
profile therefore results from the 2-D convolution
of eq.(\ref{eq:selfsimtemppredfinal}) with the \textit{WMAP} beam
profile. \citet{myers2004} and \citet{bielby2007} assumed that the source is
well resolved with respect to the \textit{WMAP} beam. Under this assumption the
full form for a 2-D convolution can be approximated by a
1-D convolution\footnote{Taking a Gaussian beam as an example, if
$\sigma_{beam}$ is much smaller than the typical scale of the cluster profile
then the $\theta$ integral in \citet{baileysparks1983}'s eq.(2) which describes
the non-radial aspect (ie: the 2-D nature) of the convolution is approximately
$2\pi$. This effectively reduces the dimensionality of the convolution, which
now takes a 1-D Gaussian form.}. However, the typical cluster sizes used in SZ
studies are of the same order as the \textit{WMAP} beams and so this
approximation can start to fail. Furthermore, for profiles such as the
self-similar model which are very centrally peaked this approximation becomes
increasingly invalid. The implementation of the PSF convolution used in this
paper is fully 2-D and does not rely on such approximations. In Section
\ref{subsection:bonamente2006res} we explore the impact of this on the
\citet{bielby2007} results.


\begin{table}
 \begin{tabular}{lccccc}

\hline\hline
 Type & $P_{0}$ & $c_{500}$ & $\gamma$ & $\alpha$ & $\beta$ \\
\hline

All:Fitted & $8.403(\frac{h}{0.7})^{-\frac{3}{2}}$ & 1.177 & 0.3081 & 1.0510 & 5.4905 \\

\hline\hline
\end{tabular}
\caption{Summary of the \textit{Planck} NFW parameters as used in eq.
(\ref{eq:nfwprofile}) and described by \citet{arnaud2010}. These are the
same parameters as used by the \textit{Planck} team, the All:Fitted set.
}
\label{tb:nfw-planckpara}
\end{table}

\section{Cross-correlation methods}
\label{sec:method}

\subsection{Stacking Positions}

The choice of coordinates to use for the radio source positions and
cluster centres is important. Scatter or an offset in the centroid used
in the cross-correlation could cause the stacked result to appear
artifically broad or induce artifacts. However, the only sample
used in this paper for which astrometric errors are appreciable are the
\textit{WMAP} derived radio-source catalogues where the astrometric error in
both longitude and latitude is 4$'$ \citep{chen2009}.
We mitigate for this effect by using the position of the corresponding matched
5GHz source, since these have negligble astrometric error $\mathcal{O}(10'')$
\citep{gregory1996, griffith1993}. We find no evidence for an offset
between the \textit{WMAP} and 5GHz sources and hence we are confident that
astrometry error will not cause new broadening of the beam. We
also note that the stacking procedure we use is dominated by the brighter
objects, which typically have better astrometry.

For the \textit{Planck} radio source catalogues we have used the
\textit{Planck} positions since these are of high astrometric quality
\citep{planckERCSCexplain}. In our sample selection we
have rejected sources which the \textit{Planck} team estimate to have
relatively high astrometric errors. The effect of this selection in the
100GHz band is to ensure that $\sigma<0'.75$ for the $S\geq1.1$Jy sample.

For the \textit{Planck} ESZ objects we have taken the \textit{Planck} estimated
positions rather than the X-ray derived positions as the cluster centres. We do
this to avoid the complications associated with rare but potentially large
offsets between the SZ and X-ray signals which are likely caused by
merging events \citep{planckearlySZ}.

\subsection{Calculation of Profiles}
\label{subsec:profilecalcmethod}

Our cross-correlation/stacking procedures for measuring both radio point
source profiles and SZ decrements are similar to those of
\citet{myers2004}, \citet{bielby2007}  and then as updated by
\citet{utaneradio}. Ultimately, we shall be stacking/cross-correlating
\textit{WMAP} data around radio source positions and cluster centres from
catalogues, particularly from the \textit{Planck} ERCSC. To estimate a temperature profile for
an individual source \textit{j} we use,

\begin{equation}
 \Delta T_{j}(\theta) = \sum_i \frac{T_{ij}(\theta) - \overline{T}_{j}}{n_{ij}(\theta)},
\label{eq:tempdeccalc}
\end{equation}

\noindent where the sum is over the pixels, denoted \textit{i}, within a
circular annulus of radius $\theta$. Here $n_{ij}$ represents the number of
pixels within the annulus and $T_{ij}$ is the temperature recorded for the pixel
\textit{i} and source \textit{j}. $\overline{T}_{j}$ is the average background
temperature which can either be estimated locally in a surrounding annulus in a
`photometric method' or globally (see \citealt{utaneradio}). These two
background estimates make no difference in the stacked results but can make a
difference for individual sources (see Section \ref{sec:szresults}). We then
stack the \textit{WMAP}7 data by averaging $\Delta T_j(\theta)$ over the sources
that have pixels within the annulus $\theta$ using the statistical average,
$\Delta T(\theta) = \sum_j (1/N_{\theta}) \Delta T_j(\theta)$. $N_{\theta}$ is
the number of sources that have pixels within the annulus $\theta$ and is
usually constant for all except the $\theta\la4'$ bins. 

We have followed \citet{utaneradio} in using jack-knife errors, for
both the radio and SZ sources, based on 6 equal area sub-fields defined by lines
of constant galactic longitude and split by the galactic equator. For $N=6$ fields
denoted \textit{k}, the errors are,

\begin{equation}
 \sigma^{2}(\theta) = \frac{N-1}{N} \sum_{k}^{N} \Big( \Delta T_{k}(\theta) - \overline{\Delta T}(\theta) \Big)^{2},
\label{eq:errdefintion}
\end{equation}

\noindent where $\Delta T_{k}(\theta)$ is the average of the fields excluding
field $k$. We have experimented with both alternative sub-fields and methods
such as bootstrap resampling finding approximately equivalent results.
In Section \ref{sec:sim} we have used simulations to test whether our method can
robustly recover the beam profile, in doing so we find that our jack-knife
errors are reasonable.



%
\section{Flux Comparisons}

\subsection{\citet{gold2011} \textit{WMAP}7 and \textit{PLANCK} ERCSC }
We first compare \textit{WMAP}7 sources at Q, V, W from \cite{gold2011} to
their counterparts in the \textit{Planck} ERCSC at 100, 70 and 44 GHz.
We also compare the \textit{Planck} fluxes in the 100GHz band to the
ground-based ATCA and IRAM source fluxes previously used by
\cite{utaneradio}. 

\begin{figure}
\begin{center}
  \includegraphics[height=13.0cm]{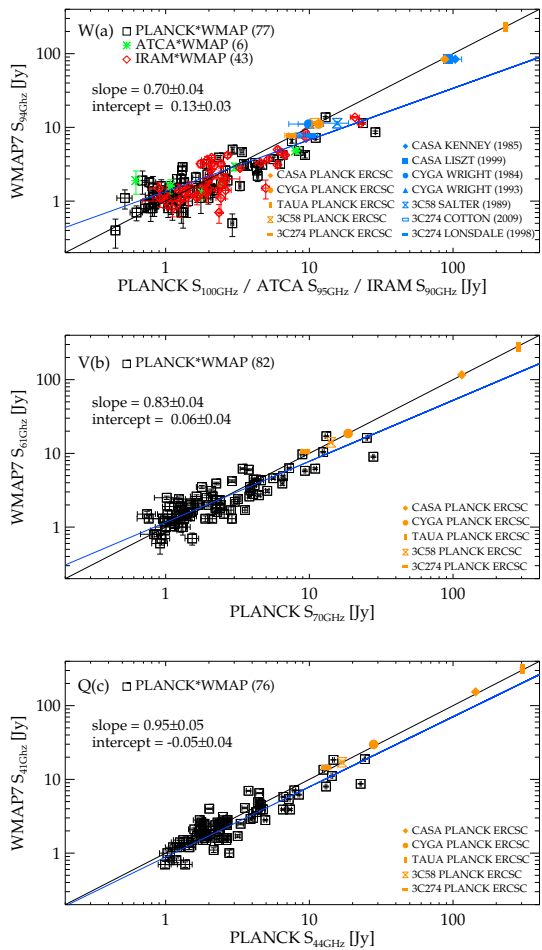}
 \caption{A comparison between the \textit{WMAP}7 fluxes, \textit{Planck} 
 and ground based source fluxes. Also shown are the one-to-one relation (black line) 
 and the best fit relation (blue line). Measurements of the \citet{weiland2011} sources
 have been corrected to a \textit{WMAP} epoch of 2005 and the respective 
 \textit{WMAP} band-centres using the \citet{weiland2011} variability estimates 
 and spectral indices.}
 \label{fig:wmapdirectfluxcompare}
\end{center}
\end{figure}

In Fig. {\ref{fig:wmapdirectfluxcompare}} we first focus on the comparison of the 
\textit{WMAP}7 fluxes to \textit{Planck} and also the ground-based ATCA and IRAM sources. 
We only consider the matches with separation less than $2'$ to avoid any
possible systematic errors associated with sources that have poor
astrometry. However, our results are independent of this cut up to separations
of $10'$. At high fluxes we see evidence for a systematically lower
\textit{WMAP} flux, $\approx 50\%$ above 2Jy. This non-linearity is
particularly prominent in the W band, the band with the greatest angular
resolution.  

Since there are errors in both variables we have used the Numerical Recipes
\citep{press1993} \textit{fitexy} as our fitting routine. We find
best fit logarithmic slopes of $[0.70\pm0.04,0.83\pm0.04,0.95\pm0.05]$ for the
[100GHz-W,70GHz-V,44GHz-Q] comparisons. To obtain realistic errors on
these fits we have linearly scaled the flux
errors until we obtained a $\chi^2$  probability of 0.5 as recommended by
\citet{press1993}, for data with a dominant intrinsic dispersion.

Clearly, \textit{Planck} and \textit{WMAP} fluxes for sources were
measured at different times. Since at least $\approx30$\% of the
\textit{WMAP}5 radio sources exhibit some level of variability
\citep{wright2009}, we expect and observe much larger scatter than
accounted for by the estimated flux uncertainty. However, we note that the
brighter \textit{WMAP} sources are fainter than the one-to-one relation, 
this is in the opposite sense expected if variability was biasing faint
\textit{Planck} sources into the \textit{WMAP} catalogue when in a
bright phase.

We investigate whether variability is alternatively causing a bias due
to \textit{Planck} dropouts by limiting the \textit{WMAP} sample to
$\geq5\sigma$ sources. The advantage is that the \textit{Planck} team have
investigated all the \textit{WMAP} $\geq5\sigma$ objects that are not in the ERCSC. They
conclude that for the 100GHz-W comparison the missing objects are `all' spurious
and can be explained by the object having a weak or missing 5GHz ID
\citep{planckERCSCexplain}. The resulting $\geq5\sigma$ \textit{WMAP} W-band
sample of 48 sources (with $S\geq0.8Jy$) is therefore complete in the
sense of being 100\% represented in the \textit{Planck} sample
with no bias due to a \textit{Planck} dropout population. When we then repeat
the \textit{WMAP}-\textit{Planck} 100GHz-W flux comparison, we find a logarithmic
slope of ($0.67\pm0.05$), consistent with the original result
and therefore strong evidence against a highly variable source population
causing dropouts that bias the \textit{WMAP}-\textit{Planck} comparison.

\begin{figure}
\begin{center}
  \includegraphics[height=6.0cm]{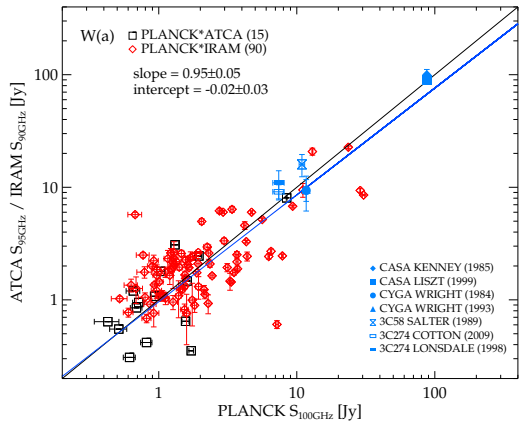}
 \caption{A comparison of the 100GHz \textit{Planck} fluxes and the ground-based
sources. Also show are the one-to-one relation (black line) and the best fit 
relation (blue line). Measurements of the \citet{weiland2011} sources
 have been corrected to a \textit{Planck} ERCSC epoch of 2010 and the respective 
 \textit{Planck} band-centres using the \citet{weiland2011} variability estimates 
 and spectral indices.}
 \label{fig:planckWdirectfluxcompare}
\end{center}
\end{figure}

The disagreement between \textit{Planck} and \textit{WMAP} is in
contrast to direct comparisons between \textit{Planck} and ground-based
ATCA/IRAM data. These instead show good agreement, as shown in Fig.
{\ref{fig:planckWdirectfluxcompare}} for the \textit{Planck} 100GHz
radio point sources. The best fit logarithmic slope of $[0.95\pm0.05]$
is statistically consistent with the one-to-one relation. However,
comparing \textit{WMAP} W-band and the ground-based ATCA/IRAM data we find
evidence for non-linearity with a best fit logarithmic slope of $[0.72\pm0.04]$.
These contrasting fits are particularly significant because the greatest
\textit{Planck}-\textit{WMAP} non-linearity comes from the 100GHz-W flux
comparison. Given the agreement between \textit{Planck} and the ground-based
ATCA/IRAM observations, we interpret the flux disagreement as being due to
\textit{WMAP}.

\subsection{Further tests for bias}

In response to the referee we have made additional bootstrap and jack-knife
re-sampling tests of the \textit{WMAP}*\textit{Planck} flux-flux comparison and its error.
After 1000 bootstrap resamplings we estimate logarithmic slopes of
$[0.70\pm0.09,0.84\pm0.11,0.96\pm0.10]$ for the
[100GHz-W,70GHz-V,44GHz-Q] comparisons. We also perform Jacknife resamplings of
the \textit{WMAP}*\textit{Planck} flux-flux comparison, we estimate a
logarithmic slope of $[0.70\pm0.10,0.83\pm0.12,0.95\pm0.10]$ for the
[100GHz-W,70GHz-V,44GHz-Q] comparisons. These resampling
results are consistent with our original samples and support the accuracy of our previous error analysis.

We have also made Monte-Carlo simulations of the flux-flux comparison.
We generated samples with the same number of sources as in the real
flux-flux comparison using the \textit{WMAP} Q-band power-law
distribution, $N(<S) \propto S^{-1.7}$, \citep{bennett2003}. These
fluxes are then scaled to the respective \textit{WMAP} and
\textit{Planck} band centres on the basis of a Gaussian distribution in
spectral indices, $\alpha$, with mean -0.09 and standard deviation
0.176, \citep{wright2009}. Realistic Gaussian measurement errors are
then assigned as a function of flux in a manner consistent with the
original \textit{WMAP} and \textit{Planck} samples. To include
variability we start from the \citet{wright2009} analysis of the
\textit{WMAP}5 data that measured  a median rms flux variability for the
25 brightest Q-band objects of $\sigma=0.23$ and which we therefore
additionally apply to all our sources, assuming a Gaussian distribution.
We then  impose detection limits corresponding to the faintest source in
the given band for the \textit{WMAP} and \textit{Planck} fluxes
respectively. Finally, we compare these two flux types by measuring the
best fit relation in the same way as was originally done for the
\textit{Planck}-\textit{WMAP} comparison. After 10,000 simulations of
the 100GHz-W comparison we find average logarithmic slopes and
intercepts of $[0.98\pm0.06]$ and $[0.04\pm0.06]$. These results are not
only in agreement with a one-to-one relation but support the errors
found in our original \textit{WMAP}-\textit{Planck} comparison. We
therefore conclude we are able to robustly recover the expected
one-to-one result and hence that our comparison may be unbiased.

\subsection{Potential contamination of \textit{Planck} fluxes by CO emission}

\citet{planckERCSCexplain} have noted that the 100GHz bandpass
contains the $J = 1\rightarrow0$ rotational CO emission line. This is a
potential explanation for the flux non-linearity we report between the
\textit{WMAP} 94GHz W-band and \textit{Planck} 100GHz bands. However, such an
explanation would imply that \textit{WMAP} and \textit{Planck} are in agreement
away from the galactic plane where CO emission is lower. However, we see no
evidence for such a distinction, with galactic latitude cuts of
$|b|>5^\circ$, $|b|>30^\circ$ and $|b|>45^\circ$ we find 100GHz-W logarithmic
slopes of $[0.70\pm0.04]$, $[0.65\pm0.06]$, and $[0.72\pm0.08]$ respectively.

\subsection{Inclusion of the 5 additional \citet{weiland2011} bright sources}

\citet{weiland2011} have made a comparison of \textit{WMAP} fluxes of bright
radio source fluxes from ground-based telescopes and claim that they support
the \textit{WMAP} flux scale. Some of the sources used are planets and have not
been through the same reduction procedures as the CMB maps but five
other sources, Cyg A, Cas A, Tau A, 3C58 and 3C274 have gone through the same
procedures. \citet{weiland2011} selected these sources on the basis that they
were the brightest and least variable of the sources with adequate background
contrast and a history of prior observation. 

We now expand our flux comparisons by including\footnote{We do not include Tau A
because it lacks a \textit{WMAP} independent measurement to compare to at the
W,V,Q frequencies.} these \citet{weiland2011} sources in Fig.
{\ref{fig:wmapdirectfluxcompare}}. We use the \textit{WMAP} fluxes quoted by
these authors and the independent ground-based fluxes that are mostly those
quoted by these authors. We see that in the W band at least, \textit{WMAP} also
underestimates the flux of these sources (blue points) and indeed Cyg A,
3C274 and 3C58 appear not inconsistent with our fitted relation. However, the
underestimation for Cas A is less than predicted by extrapolating the fit to the
brighter radio fluxes. If this result were to be more highly weighted then there
would still be evidence for a \textit{WMAP} flux problem, but one which now
looked more like a constant offset than a scale error. However, we note that
there are differences between the two ground-based observations of Cas A.
Furthermore, Cas A lies close to the galactic plane ($|b|<6^\circ$) and hence
contamination might be an issue. It may therefore be too early to infer a flux
offset on the basis of this source.

When we include the Celestial sources from \citet{weiland2011} with independent
ground-based measurements \footnote{We do not include measurements without error
estimates, this excludes the Cyg A \citet{wright_sault1993} and 3C274
\citet{lonsdale1998} measurements} into the \textit{Planck}-\textit{WMAP}
comparison we find logarithmic slopes of $[0.86\pm0.08,0.90\pm0.10,0.98\pm0.09]$
for the [100GHz-W,70GHz-V,44GHz-Q] comparisons. After Jack-knife and bootstrap
resampling we find logarithmic slopes of $[0.81\pm0.10,0.83\pm0.12,0.95\pm0.10]$
and $[0.81\pm0.08,0.84\pm0.11,0.986\pm0.10]$ respectively  for the
[100GHz-W,70GHz-V,44GHz-Q] comparisons. Hence, whilst including
the Celestial source data changes the degree of the non-linearity, the results
are still in significant disagreement with a one-to-one relation.

We finally add the \textit{Planck} ERCSC measurements of these 5 sources to Fig.
{\ref{fig:wmapdirectfluxcompare}} and Fig. {\ref{fig:planckWdirectfluxcompare}}.
This complicates the picture further since they appear to agree with the
\textit{WMAP} results more than the ground-based results. As far as we can
see, the \textit{Planck} fluxes are not calibrated via \textit{WMAP}. If we then
fit the full \textit{Planck}-\textit{WMAP} W band comparison we now find less
evidence for a discrepancy between the two finding a 100GHz-W logarithmic
slope of $[0.91\pm0.04]$. But just making the 100GHz-W comparison in the 3-400Jy region,
the result might then again look more like a constant offset with a logarithmic
slope and intercept of $[1.01\pm0.10]$ and $[0.16\pm0.12]$.

We conclude that the \textit{WMAP} fluxes in the $S\approx10$Jy region show
systematically lower fluxes than \textit{Planck} and we have argued that this
discrepancy is unlikely to be explained by variability, underestimated
errors or inaccurate flux estimation. At lower and higher fluxes the
\textit{WMAP}-\textit{Planck} agreement seems better, implying some
non-linearity in their relative scales. If \textit{WMAP} data are compared  to
ground-based data rather than \textit{Planck}, the same discrepancy is seen at
$S\approx10$Jy and a small but significant discrepancy is seen at brighter
fluxes, which would more imply a linear offset rather than a non-linearity.
Similar effects are seen at Q and V but at a lower level.

\section{\textit{WMAP} point source profiles}
\label{subsec:planckradioresults}

\subsection{\textit{Planck} ERCSC radio sources}

\begin{figure*}
 \includegraphics[width=17cm]{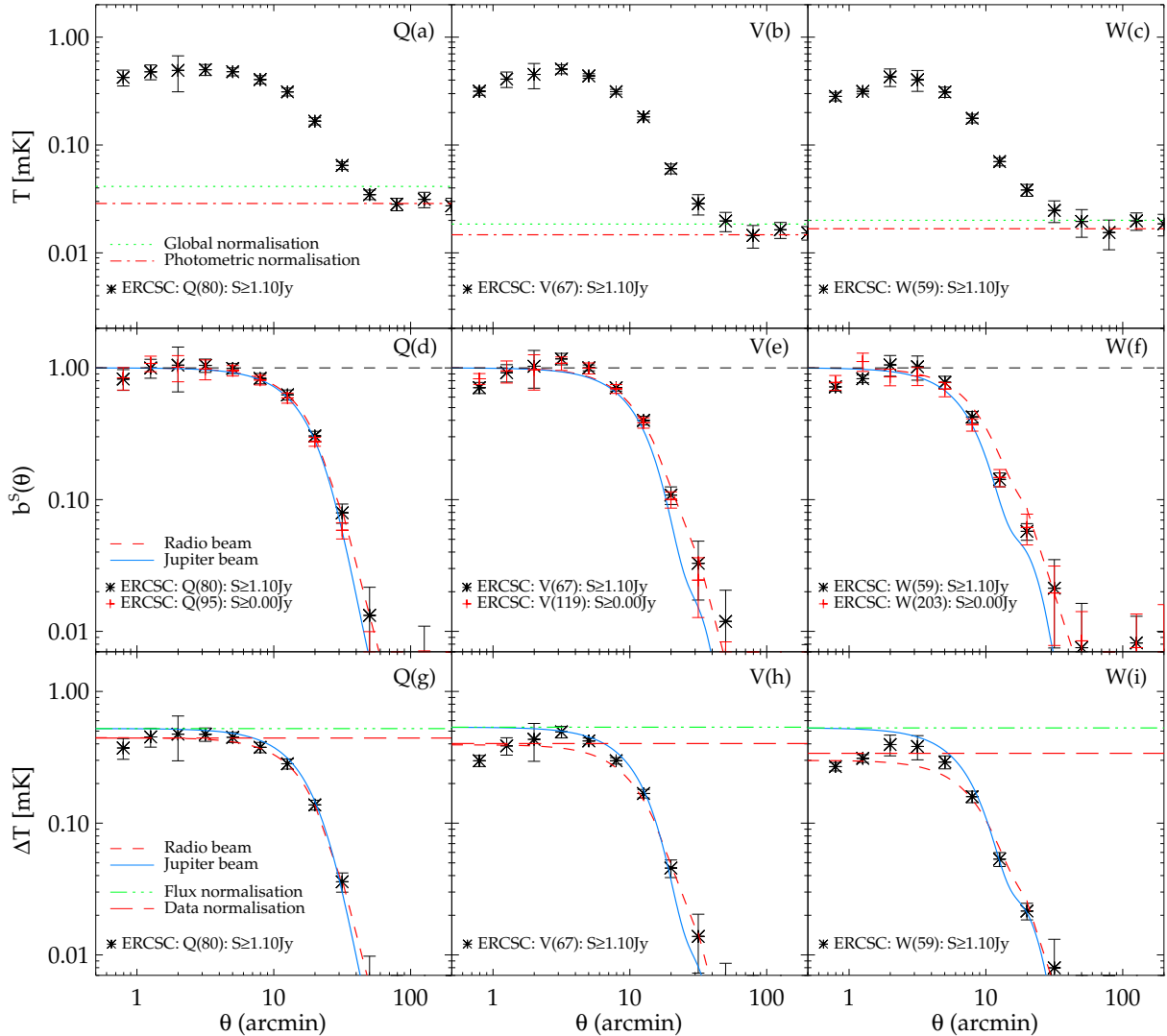} 
\caption[]{(a),(b),(c): The raw stacked \textit{WMAP}7 [Q,V,W] temperature
profiles for the \textit{Planck} [44,70,100] GHz band sources with the
global mean and photometric background temperatures of the map plotted as dashed (green, red) lines.
(d),(e),(f): The photometrically subtracted, stacked and re-normalised \textit{WMAP}7 [Q,V,W] $b^{s}(\theta)$
profiles for the \textit{Planck} [44,70,100] GHz band sources. Also shown are
the $b^{s}(\theta)$ for the Jupiter beam (blue, solid) and the radio source fit (red, dashed)
of \citet{utaneradio}. Any sensitivity to Eddington bias is shown in the
profiles without the flux limit of $S\geq1.1$Jy (lighter red, cross).
(g),(h),(i): The photometrically subtracted and stacked  \textit{WMAP}7  [Q,V,W] $\Delta T(\theta)$
profiles for the \textit{Planck} sources. 
Also shown are the $\Delta T(\theta)$ for the Jupiter beam (blue, solid)  and the radio source fit (red, dashed)
of \citet{utaneradio}, now absolutely normalised via the \textit{Planck} flux.
}
\label{fig:radioproftempandbsplanck}
\end{figure*}

We now apply the stacking analysis of \citet{utaneradio} to the
co-added \textit{WMAP}7 maps, centring on the \textit{Planck} radio point sources.
\textit{Planck} sources are selected at multiple wavebands which may be
advantageous in avoiding spurious sources etc. Figs.
\ref{fig:radioproftempandbsplanck} (a),(b),(c) are shown for
completeness because these raw temperature plots demonstrate the main
uncertainty in this analysis which is the accuracy of the background
subtraction. We note that there is some difference between the global
background and the background local to the source samples
but generally this effect appears smaller in the \textit{WMAP}7 data (eg
at W) than it was in the \textit{WMAP}5 datasets used by
\citet{utaneradio}.

Figs. \ref{fig:radioproftempandbsplanck} (d),(e),(f) show the same profiles
now background subtracted and scaled  to unity at the origin  to produce
$b^{S}(\theta)$. We have used the `photometric' subtraction to produce
the radio point source temperature profile, $\Delta T_{radio}(\theta)$.
For the \textit{WMAP}7 dataset there is very little difference in the profiles
resulting from global or local/photometric background subtractions.

These $b^{S}(\theta)$ are now compared to the \textit{WMAP} Jupiter beam
and the best fit beam to the bright \textit{WMAP} radio source profiles
found by \citet{utaneradio} (dashed orange line in their Fig. 2). There
is again evidence that the \textit{Planck} selected radio sources
suggest a wider beam than the Jupiter beam, particularly in the W band,
although the \textit{Planck} sources lie slightly below the profile fits
from \citet{utaneradio}. We further note that the statistical significance of the
deviations from the Jupiter beam for  the \textit{Planck} selected sources at
$12'.6-19'.9$ is only modest at $\approx2-3\sigma$ for the W band.

The normalisation of $b^{S}(\theta)$  to unity at small scales forms a
further uncertainty in these beam comparisons. In Figs.
\ref{fig:radioproftempandbsplanck} (g),(h),(i) we have applied the formalism
of Section \ref{subsec:model-radiofluxes} and attempted to make  absolute
normalisations of the various model profiles, using the \textit{Planck} ERCSC
listed fluxes. We assume in turn the Jupiter profile and then the radio source
profile of \citet{utaneradio} in calculating the resulting $\Gamma^{ff}$ factor.
These give respectively the blue and red lines. Hence, if the radio sources
followed the Jupiter profile, for example, we should see the same peak
temperature for the  stacked model profile and the stacked data. We see that the
\textit{Planck} peak temperatures, particularly in the W band, tend to lie
between the Jupiter profile and the previous \textit{WMAP} bright radio source
fits. These results suggest that the previous radio source fit may be too wide
at $\theta>30'$ where it is essentially an extrapolation, unconstrained by
the data, and this will affect the accuracy of its absolute
normalisation i.e. there is a large error in $\Omega_{\rm beam}$.  
Otherwise, the conclusion is similar to that from Figs.
\ref{fig:radioproftempandbsplanck} (d),(e),(f) in that  the \textit{Planck}
data is suggesting that the Jupiter beam is a poor fit to the radio
source profiles particularly at W.

The question of Eddington bias was discussed by \cite{utaneradio} and
has also been suggested by \cite{schultz2011} as a possible explanation
of the wide radio profiles. In terms of the \textit{Planck} sources an
Eddington bias of $\approx$0.02mK is required to explain our results. However
our pre-selection of these sources as being point-like at \textit{Planck}
resolution and our rejection of both faint ($S<1.1$Jy) and CMB-contaminated
sources mean that it is difficult to see how Eddington bias could be affecting
these results. In Fig. \ref{fig:radioproftempandbsplanck} (d),(e),(f) we have
also presented the source sample without the $S\geq1.1$Jy flux cut. The
consistency of the full source and brighter source samples indicates that
Eddington bias is not significantly affecting these samples.

\begin{figure*}
 \includegraphics[width=17cm]{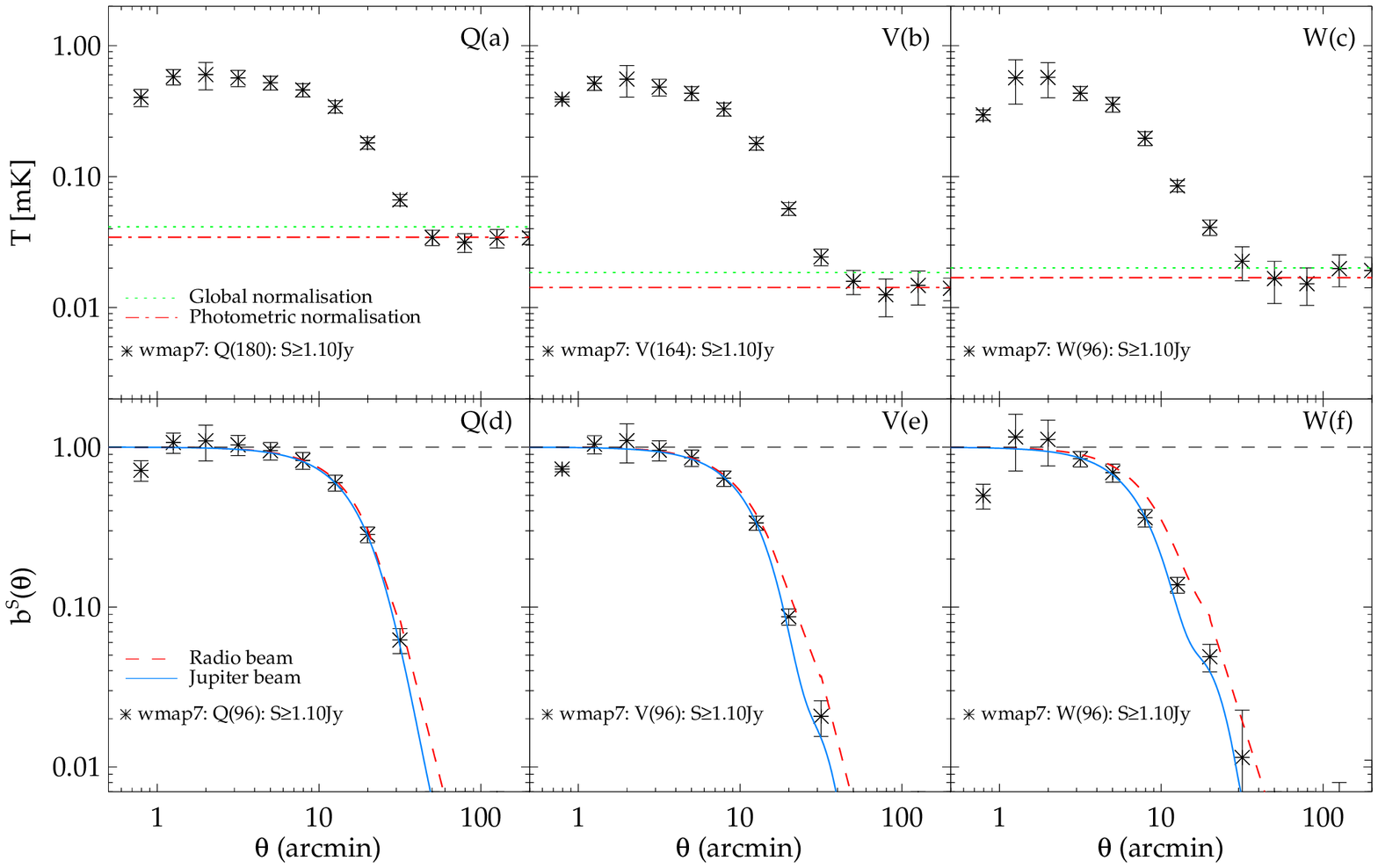} 
\caption[]{(a),(b),(c): The raw stacked \textit{WMAP}7 [Q,V,W]
temperature profiles for the \textit{WMAP}7 sources of \citet{gold2011}
with the global mean and photometric background temperatures of the map plotted as dashed (green, red) lines.
(d),(e),(f): The photometrically subtracted, stacked and re-normalised
\textit{WMAP}7 [Q,V,W] $b^{s}(\theta)$ profiles for the \textit{WMAP}7
sources of \citet{gold2011}. Also shown are the $b^{s}(\theta)$ for the
Jupiter beam (blue, solid) and the radio source fit (red, dashed) of
\citet{utaneradio}.
}
\label{fig:radioproftempandrawwmap} 
\end{figure*}

\begin{figure*}
 \includegraphics[width=17cm]{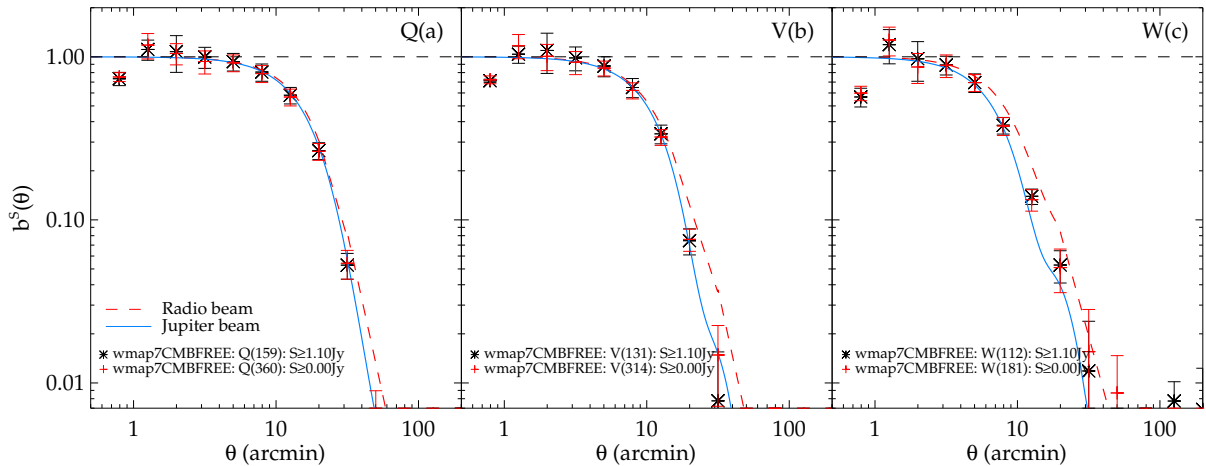}
\caption[]{(a),(b),(c): The photometrically subtracted, stacked and re-normalised \textit{WMAP}7
[Q,V,W] $b^{s}(\theta)$ profiles for  the CMB-free \textit{WMAP}7
catalogues of  \citet{gold2011}. Also shown are the
$b^{s}(\theta)$ for the Jupiter beam (blue, solid) and the radio source fit
(red, dashed) of \citet{utaneradio}. Any sensitivity to Eddington bias is shown
in the profiles without the flux limit of $S\geq1.1$Jy (lighter red, cross).
}
 \label{fig:radioproftempandbscmbfree}
\end{figure*}

\begin{figure*}
 \includegraphics[width=17cm]{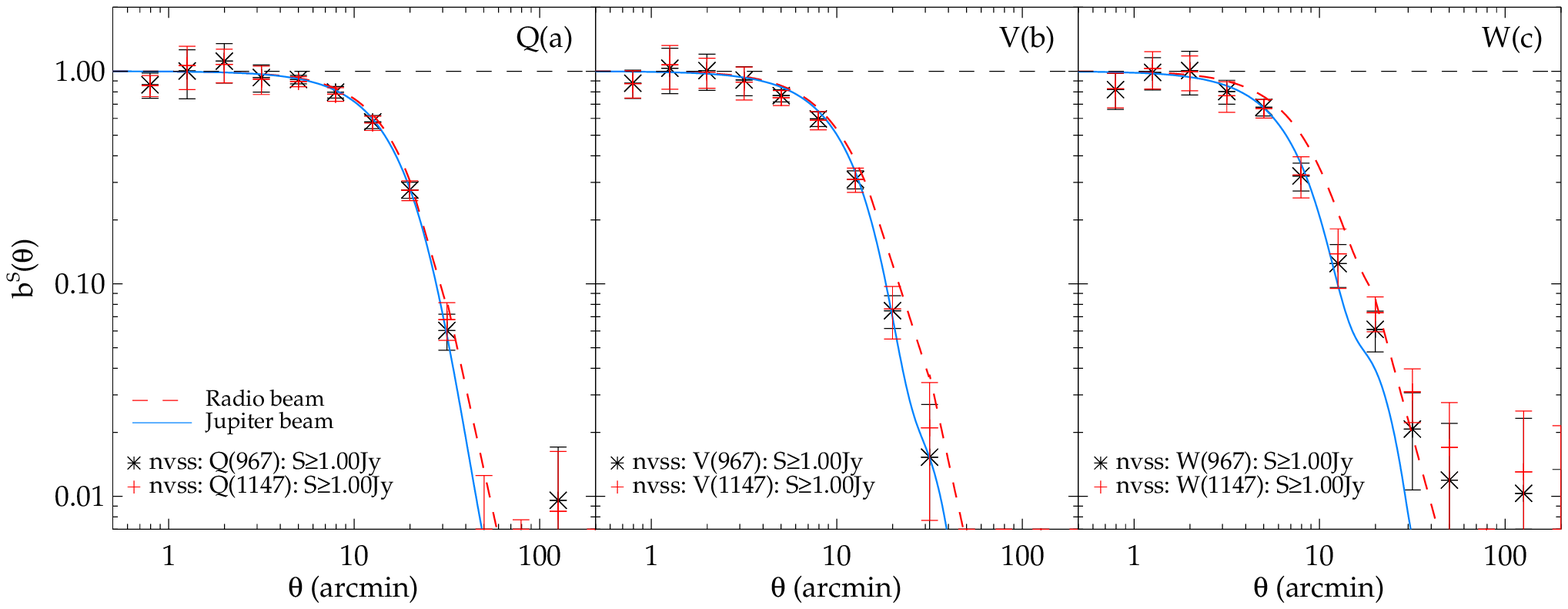}
 \caption[]{(a),(b),(c): The photometrically subtracted stacked \textit{WMAP}7
[Q,V,W] $b^{s}(\theta)$ profiles for the NVSS sources. Also shown are the
$b^{s}(\theta)$ for the Jupiter beam (blue, solid) and the radio source fit
(red, dashed) of \citet{utaneradio}.}
 \label{fig:radioproftempandbsnvss}
\end{figure*}

\subsection{\textit{WMAP}7 radio sources}

We next repeat the analysis of \citet{utaneradio} using the co-added
\textit{WMAP}7 maps and source catalogue \citep{gold2011}. The results are shown
in Fig. \ref{fig:radioproftempandrawwmap}. We see that the results again tend to
lie between the Jupiter profile and the previous \textit{WMAP} bright radio
source fits by \citet{utaneradio}. This may in part be due to the
\textit{WMAP}7 profiles returning to zero at large scales more uniformly than
\textit{WMAP}5, making the differences between the photometric and
global profile estimates more marginal. However, we also found that using the
\textit{WMAP}5/\textit{WMAP}7 catalogue in the \textit{WMAP}7/\textit{WMAP}5
temperature maps gives profiles more consistent with the \citet{utaneradio}
fits. We therefore attribute the difference between Fig.
\ref{fig:radioproftempandrawwmap} and the results of \citet{utaneradio} to a
possible systematic difference between \textit{WMAP}5 and
\textit{WMAP}7, with perhaps a contribution from statistical fluctuations.

Following \citet{utaneradio} we minimised any effect of Eddington bias for this
sample by pre-selecting only sources that appear in the 5GHz GB6 and PMN radio
samples. We have only used 5GHz coordinates for the radio sources, with their
sub-30$''$ accuracy to minimise any positional error in our analysis. Although
Eddington bias may well be affecting the faintest \textit{WMAP} sources, as was
also noted in \citet{utaneradio}, we have used a flux limit of $S\geq1.1$Jy. In
Section \ref{sec:sim} we shall check for the presence of Eddington bias in this
particular dataset using random simulations.

\subsection{\textit{WMAP}7-CMBfree radio sources}

In the `CMB-free' method \citep{chen2009},\textit{WMAP} sources are
selected using the Q,V,W bands simultaneously to form an internal linear
combination map (ILC) with weights chosen to cancel out the CMB
anisotropy signal. Again, any Eddington bias due to CMB fluctuations
should be reduced in the case of this point source catalogue. We
therefore repeated our stacking analysis with the 417 QVW sources from
the \citealt{gold2011} \textit{WMAP}7 `CMB-free' catalogue (see Fig.
\ref{fig:radioproftempandbscmbfree}). Overall we again see
wider-than-expected profiles at W, broadly consistent with the results
in Figs. \ref{fig:radioproftempandbsplanck},
\ref{fig:radioproftempandrawwmap}. Finally, we have also presented these results
without the $S\geq1.1$Jy flux limit, we note that the result is unchanged. This 
consistency is evidence for robustness of the result to Eddington bias.

\subsection{NVSS radio sources}

Point source catalogues made at significantly lower frequencies than the
\textit{WMAP} bands are unlikely to be affected by Eddington bias
due to CMB fluctuations, if identification is done independently of the
\textit{WMAP}7 point source catalogue. For example, point sources
selected at 1.4 GHz will have Rayleigh-Jeans temperature $\approx
4500\times$ higher than a source with similar flux density selected at
W-band ($\approx 94$ GHz), i.e. $T_{\rm RJ} \propto \Omega_{\rm
beam}^{-1}\nu^{-2}$, whereas the rms Rayleigh-Jeans temperature due to
the CMB fluctuations stays roughly the same between the two frequency
bands, \citep{bennett2003}. Therefore, we now stack co-added \textit{WMAP}7
temperature data centred around the positions of the 1147 $S_{1.4} > 1$
Jy NVSS point sources. Fig. \ref{fig:radioproftempandbsnvss} shows the
resulting Q, V and W profiles. We see that they are consistent with
those measured using \textit{WMAP}5 total/bright sources in Fig. 2 of
\citet{utaneradio}. However, the profiles do not appear as wide as the
\textit{WMAP}5 faintest subsample despite the average flux of the
NVSS sample at \textit{WMAP} bands being $\approx 3\times$ lower.

Many of the NVSS sources are resolved into multiple components
\citep{blake2002}. However, this is unlikely to cause the widening of
the beam beyond $\theta \ga 6'$. Here, as a precautionary measure, we
shall test the beam profile measured using the NVSS by excluding any
source that has neighbouring source(s) within $1^\circ$. This extra
condition reduces the number of $S_{1.4} > 1.0$ Jy sources outside the
\textit{WMAP}7 `point source catalogue' mask to 967. The resulting co-added 
beam profiles are also shown in Fig. \ref{fig:radioproftempandbsnvss}. We see
that the beam profiles are in good agreement with the previous results.

\section{\textit{WMAP} and NVSS source catalogue simulations}
\subsection{Description}
\label{sec:sim}

We made 100 Monte Carlo simulations to check our method and the
robustness of the results. These simulations are due to Sawangwit (2011)
who made them in the context of his test of the W1 beam
in the \textit{WMAP}5 dataset. Thus they are conservative in terms of 
both the errors they imply and the test of Eddington bias they make 
in our new context of the averaged DA's (W1-W4 in the W-band case)
and the \textit{WMAP}7 dataset. We followed the procedures described by
\cite{wright2009} \citep[see also][]{chen2009}. For each set of
simulations, $\approx10^6$ point sources are generated with a power-law
distribution, $N(>S)\propto S^{-1.7}$, at \textit{WMAP} Q-band
\citep{bennett2003,chen2009}. Their spectral indices, $\alpha$, are
drawn from a Gaussian distribution with a mean -0.09 and standard
deviation 0.176 as characterised by the \textit{WMAP}5 point source
catalogue \citep{wright2009}. The flux density for each object is scaled
to the centre of the other four bands using the relation $S_{\nu}
\propto \nu^{\alpha}$. The source positions are then randomly
distributed on the sky and each source is assigned to a pixel in a
HEALPix $N_{side}=2048$ map. For a source with flux density $S_{\nu}$,
the peak Rayleigh-Jeans temperature difference, $\Delta T_{RJ}(0)$, is
given by eqs.({\ref{eq:gammatemprel}}, {\ref{eq:gammadef}}), but with the
$\Omega_{beam}$ replaced by $\Omega_{pix}=2.5 \times 10^{-7}$ sr, solid
angle of a $N_{side}=2048$ pixel. The publicly available \textit{WMAP}
maps (Section \ref{sec:wmap_data}) are given in thermodynamic temperature
\citep{Limon08}. For a direct comparison with our results, we thus
convert the simulated source's $\Delta T_{RJ}(0)$ to $\Delta T_{t}(0)$
using eq.({\ref{eq:anttothermo}}).

Five temperature maps, one for each band, are then smoothed with the
corresponding \textit{WMAP} beam transfer function \citep{hill2009}
before being downgraded to $N_{side}=512$. The simulated CMB temperature
map (smoothed with an appropriate beam transfer function) constructed
from \textit{WMAP}5 best-fit $C_\ell$ and pixel noise are then added to the
source temperature maps. The pixel noise is modelled as a Gaussian
distribution with zero mean and standard deviation
$\sigma=\sigma_0/\sqrt{N_{\rm obs}}$, where $N_{\rm obs}$ is the number
of observations in each pixel and $\sigma_0$ is given for each DA and
frequency band \citep{Limon08}. Here, we use the \textit{WMAP}5 $N_{\rm
obs}$ map to generate pixel noise for its corresponding band map.

\subsection{Source Detection}

Next, we applied the five-band detection technique following procedures 
utilised by \textit{WMAP} team \citep{bennett2003,gold2011}. 
Firstly, the temperature maps are weighted by the number of observations 
in each pixel, $N_{\rm obs}^{1/2}$. The weighted map is then filtered in 
harmonic space by $b_\ell/(b^2_\ell C_\ell^{\rm CMB}+C_\ell^{\rm noise})$ 
\citep[e.g.][]{Tegmark98,refregier2000} 
where $C_\ell^{\rm CMB}$ is the CMB power spectrum and $C_\ell^{\rm
noise}$ is the noise power, and $b_\ell$ is the beam transfer function
\citep{hill2009}. The filter is designed to suppress fluctuations due to
the CMB at large scales and pixel noise at scales smaller than the beam
width. We used the \textit{WMAP}5 best-fit $C_\ell$ for $C_\ell^{\rm CMB}$.
The $C_\ell^{\rm noise}$ are determined from pixel noise maps
constructed using $\sigma_0$ and five-year $N_{\rm obs}$ for each band
as described above. We then search the filtered maps for peaks which are
$> 5\sigma$. Peaks detected in any band are fitted to a Gaussian profile
plus a planar baseline in the unfiltered maps for all other bands. The
recovered source positions are set to the best-fit Gaussian centres in
W-band. The best-fit Gaussian amplitude is converted to Rayleigh-Jeans
temperature, using the relation given in eq.({\ref{eq:anttothermo}}), and
then to a flux density using conversion factors, $\Gamma^{ff}(\nu)$,
given in Table 4 of \cite{hill2009}. In any given band, we only use
sources that are $>2\sigma$ and the fitted source width smaller than 2x
the beamwidth, following the \textit{WMAP} team. The number of detected
sources from 100 realisations are consistent with \textit{WMAP}5 point
source analyses by \cite{wright2009} and \cite{chen2009}. Our
simulations also recover the input power-law $N(>S)$ distribution down
to the expected \textit{WMAP}5 limit, $S\approx1$ Jy, remarkably well
\citep[see][]{SawangwitThesis}.

\begin{figure*}
\centering
\includegraphics[scale=0.5]{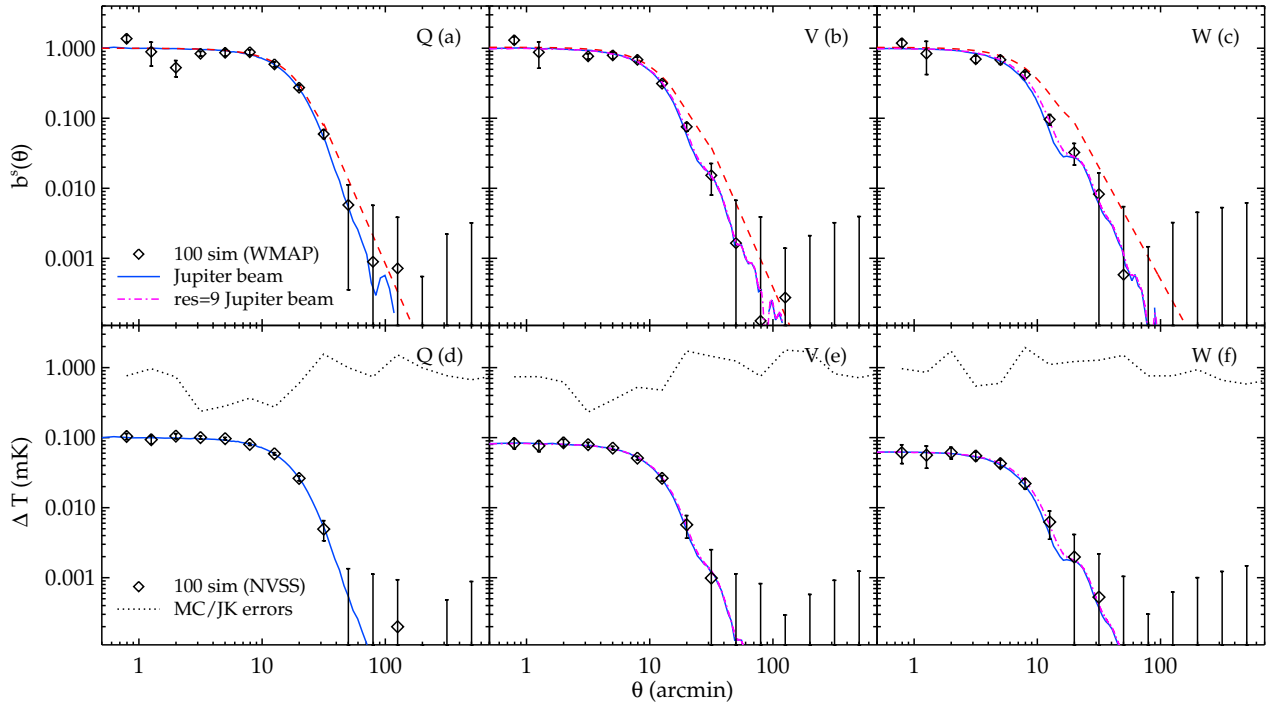}
\caption[The Monte Carlo simulation results for the beam profiles measured using NVSS sources, 
comparing to the real measurements]
{(a),(b),(c): The recovered [Q1,V1,W1] beam profiles using simulated \textit{WMAP} point sources. The error bars are 1$\sigma$ 
rms of 100 simulations. The effect of pixelisation on the profile measurement is shown by the magenta dot-dashed 
lines. (d),(e),(f): Similar to (a-c) but now using $S_{1.4} > 1$ Jy NVSS sources and without the re-normalisation. 
The ratios between Monte Carlo and jack-knife errors are shown as the dotted lines.}
\label{fig:sim}
\end{figure*}


\subsection{\textit{WMAP} simulation results}

For each simulation we applied our beam profile analysis outlined in Section
\ref{sec:method} (including a flux cut of $S>1.1$Jy). The average beam profiles derived from 100 simulations
are plotted in Fig. \,\ref{fig:sim}(a)-(c) where the error bar represents
their standard deviation in each angular bin. We found that even
profiles as narrow as the W1-band Jupiter profile can be retrieved
remarkably well out to $30'$. The estimated uncertainties using these
Monte Carlo simulations are consistent with the jack-knife error
estimations. Note that the Monte Carlo error converges after
$\approx$60-70 simulations. The Monte Carlo simulations we performed
here suggests that our method for recovering beam profile by stacking
temperature maps around point sources is robust and the jack-knife error
estimation is reliable.

\subsection{NVSS simulation results}

Although we argued above that sources (i.e. their identifications and positions)
selected at NVSS frequency are robust against the CMB fluctuations compared to
\textit{WMAP} bands, our beam analysis is still carried out using \textit{WMAP}
temperature maps. As we noted, the average flux of the $S_{1.4} > 1$Jy NVSS
sources in the \textit{WMAP} bands is $\approx 3$x lower than the \textit{WMAP}
sample. Therefore it is important to check whether the \textit{WMAP} beam
profiles can be robustly recovered using these NVSS sources. Again the results
come from Sawangwit (2011) and were only applied to the W1 detector assembly and
use \textit{WMAP}5 data.
  
We again created 100 Monte Carlo simulations similar to those described
above but without the five-band detection procedure since these sources
are pre-detected by NVSS with high positional  accuracy ($\la 1''$,
\citealt{Condon98}). The 967 NVSS source positions are used and fluxes
at 1.4 GHz are extrapolated to \textit{WMAP} Q, V and W bands assuming a
mean spectral index, $\alpha$, of -0.45 in order to mimic the average
flux density observed in these bands. The temperature maps are smoothed
with the corresponding \textit{WMAP} (Jupiter) beam profiles. The
simulated CMB fluctuations and radiometer noise are then added to the
source temperature maps as described above. For each \textit{WMAP} band,
we applied our beam profile analysis to each of the 100 simulated maps (including a flux cut of $S_{1.4}>1$Jy).
The results are shown in Fig. \ref{fig:sim}(d)-(f). The plot shows that
with these NVSS radio sources the \textit{WMAP} beam profiles can be
robustly recovered out to $30'$ and are not affected by the source
clustering consistent with the semi-empirical calculation presented in
\cite{utaneradio}. We then take the standard deviation of the 100
simulated results in each angular bin as the $1\sigma$ error. The ratio
of the Monte Carlo error to the jack-knife error is shown as the dotted
line in Fig. \ref{fig:sim}(d)-(f). The Monte Carlo and jack-knife errors
are in good agreement except at scales $< 10'$ where jack-knife errors
are somewhat over-estimates in Q and V bands.

The simulations suggest that when flux limited at $S\geq1.1Jy$, the \textit{WMAP} 
selected source profiles are unaffected by Eddington bias. The simulations also 
support the accuracy of our empirical errors. The simulations suggest 
the same conclusions apply when dealing with flux-limited ($S_{1.4} > 1$Jy) NVSS data.

\section{Possible explanations of wide radio source profiles}
\label{sec:flux_nl}

\begin{figure}
\begin{center}
  \includegraphics[width=7.5cm]{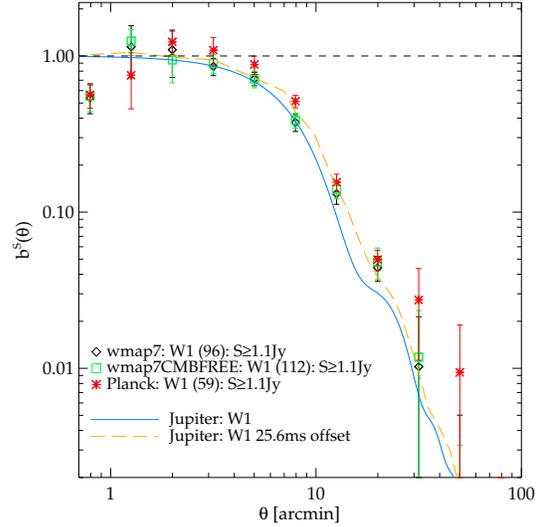}
 \caption[]{The photometrically subtracted, stacked and re-normalised \textit{WMAP}7
[Q1,V1,W1] $b^{s}(\theta)$ profiles for the \textit{Planck}, \textit{WMAP}7 and CMB-free \textit{WMAP}7
catalogues. Also shown are the $b^{s}(\theta)$ for the Jupiter beam (blue, solid) and a model showing 
the effect of a 25.6ms timing offset on the W1 Jupiter beam (orange, dashed)
\citet{SawangwitThesis}.}
\end{center}
\label{fig:W1_offset_beam}
\end{figure}


We now briefly consider possible explanations for the radio source
profiles we have observed. We start by accepting that in the
\textit{WMAP}7 data the profiles are less wide than found in the
\textit{WMAP}5 data discussed by \citet{utaneradio}. We regard our best
current result to come from comparison with the \textit{Planck} radio
sources where we have checked the sources against spatial extension at
\textit{Planck} resolution and also rejected any that are contaminated
by the CMB. This sample still rejects the W-band Jupiter beam at
$\approx2-3\sigma$ significance at $12'.6-19'.9$, about the same as the
rejection of the previous best fit model from \citet{utaneradio}. Therefore it
is not outside the bounds of possibility that the previous result may be
explained by a statistical fluctuation in the \textit{WMAP}5 data.
However, at the suggestion of the referee, we have now evaluated the
rejection significance of the Jupiter beam using the full covariance
matrix from our simulations, fitting in the range $4'<\theta<20'$. For
the Jupiter beam in the 7yr coadded maps we find formal rejection significances
from the $\chi^2$ distribution of [$1.5^{-3}$, $4.5^{-4}$, $1.2^{-3}$,
$1.4^{-6}$] for \textit{Planck}, \textit{WMAP}7, \textit{WMAP}7-CMBfree and NVSS
respectively. Although we note that the overlap between these samples means that
these results cannot be simply combined, individually these probabilities
correspond to $\ga3\sigma$ rejections of the Jupiter beam.

It is therefore still worth considering whether a wider beam could be related
to other possible \textit{WMAP} data problems. The first of these is the
possible disagreement in \textit{WMAP} flux comparisons with ground-based and
\textit{Planck} datasets noted by \citet{utaneradio} and also in this paper.
Certainly a non-linearity like we first fitted to Fig.
\ref{fig:wmapdirectfluxcompare} goes in the right direction to explain a flatter
than expected profile. Indeed, if the addition of Cas A, Cyg A, Tau A, 3C274 and
3C58 does imply that \textit{WMAP} fluxes are simply offset from \textit{Planck}
and ground-based fluxes, then flux comparisons would be consistent with the
wide beam. A logarithmic intercept of $\approx -0.1$ implies the
\textit{WMAP} flux is $\approx$80\% of the corresponding \textit{Planck}
flux. Equally, we find the W-band \textit{WMAP} Jupiter
beam solid angle is $\approx$80\% of the 25.6 ms timing offset
derived beam's $\Omega_{beam}$. This is in agreement with the expectation
from eq.(\ref{eq:gammatemprel}) that at fixed temperatures (ie: those
provided in the \textit{WMAP} maps) $S_{tot} \propto \Omega_{beam}$.

However, more data at brighter fluxes is needed to check if the
\textit{WMAP} flux is non-linear or simply offset with respect to other
datasets. We note that \citet{malik11} has used the CMB dipole to look for
non-linearity in the \textit{WMAP} temperature scale and failed to find any
evidence for such an effect.

The second possible explanation for the wider than expected radio source
profiles focused on the possibility that there was a timing offset
between the \textit{WMAP} antenna pointing and temperature data, as
proposed by \citet{liuli}. As well as causing effects at large scale due
to a wrongly subtracted dipole, this scan pattern offset would cause a
wider beam profile (see \citealt{moss2011}). \citet{SawangwitThesis} have
calculated the beam pattern that a 25.6 ms timing offset would cause in
the W band. The calculation assumes the W1 Jupiter beam and takes into
account its initial asymmetry on the sky. After creating simulated \textit{WMAP}
TOD that include  point sources distributed in ecliptic latitude and
then applying mapmaking to these data, \citet{SawangwitThesis} found the
azimuth averaged beam profiles shown in Fig. \ref{fig:W1_offset_beam} for both
zero timing offset and the 25.6ms timing offset with the latter giving a
reasonable fit to the \textit{Planck} data. More details are presented by
\citet{SawangwitThesis}. These include further results based on using the
dependence
of beam shape with ecliptic latitude to try and determine the timing offset
which marginally prefer zero timing offset. We note that \citet{roukema2011}
made similar tests based on bright \textit{WMAP} sources and found no evidence
for a timing offset at the map-making stage. On the other hand, \citet{liuli} 
checked between offsets by minimising dipole residuals and found strong 
evidence for a non-zero offset, see Sawangwit et al (in prep). We note that the
\textit{WMAP} team have indicated that they use a timing offset of zero in which
case the above agreement would simply represent a coincidence.


\section{SZ Results}
\label{sec:szresults}

\subsection{\textit{Planck} Intermediate Results}
\label{subsec:planckintermediateres}

Our final aim is to make a comparison between the \textit{Planck} ESZ and
\textit{WMAP} SZ results as described in Section
\ref{sec:model-sz}, \ref{subsec:model_sz_recon} and
\ref{subsec:model_sz_convol}. However this involves reverse engineering the
\textit{Planck} SZ $\Delta T(\theta)$ profiles. We therefore first use
recently released \textit{Planck} SZ data to check our reverse engineered
\textit{Planck} profiles. A series of papers have been released as a follow up
to the \textit{Planck} ESZ data. Two papers in particular are relevant to
corroborating the \textit{Planck} profiles presented in this paper. In
\citet{mozzatta2012} the \textit{Planck} Coma SZ profile has been published.
Additionally, the `physical' \textit{Planck} SZ temperature profiles for the 62
local clusters (LSZ) in the \citet{planckszstudy2011} analysis have been
published in \citet{pointecouteau2012}. Below, we compare our reverse
engineered profiles to these \textit{Planck} data.

\subsubsection{\citet{mozzatta2012}}
\label{subsubsec:mozzatta2012}

\begin{figure}
\begin{center}
  \includegraphics[width=7.5cm]{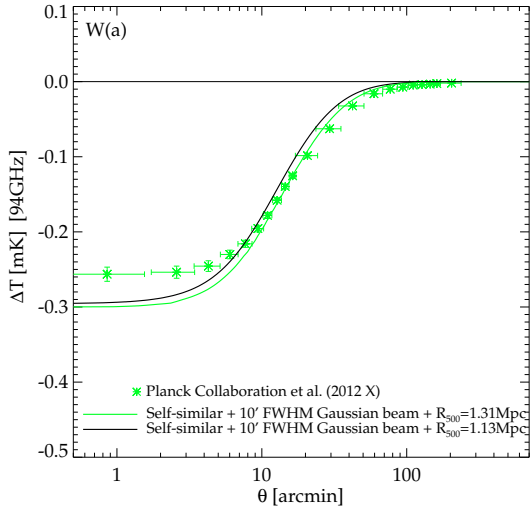}
\caption[]{(a): 
The \citet{mozzatta2012} \textit{Planck} SZ decrement for the Coma
cluster converted from RJ temperatures to a thermodynamic temperature at 94GHz.
Also shown are the \textit{Planck} temperature decrements from
eq.(\ref{eq:selfsimtemppredfinal}) using the ESZ value of $R_{500}$ (black,
solid) and eq.(\ref{eq:finalYscale}) using the \citet{mozzatta2012}
value of $R_{500}$ (green, solid). Both models are convolved with a $10'$ FWHM
Gaussian beam appropriate to the \citet{mozzatta2012} data.
}
 \label{fig:mazzotta_planckdata_coma_results}
\end{center}
\end{figure}

In Fig. \ref{fig:mazzotta_planckdata_coma_results} we now compare our
\textit{Planck} Coma SZ profiles to \citet{mozzatta2012}. We have shown two
alternative \textit{Planck} models in order to display the sensitivity of our
results to the cluster size estimates. This is motivated by the significant
difference between the value of $R_{500}=1.31Mpc$ used in \citet{mozzatta2012}
and the ESZ value, $R_{500}=1.13Mpc$. Since the value for the integrated SZ
signal, $Y(5R_{500})$, is degenerate with cluster size the ESZ value for
$Y(5R_{500})$ cannot be assumed. Therefore, in the first instance we do show an
expected \textit{Planck} temperature decrement using the ESZ values and
calculated using eq.(\ref{eq:selfsimtemppredfinal}). However, we also show a
model which uses an alternative method for calculating the expected
\textit{Planck} profile. Here, the \citet{mozzatta2012} value of $R_{500}$ is
used to calculate $Y_{500}$ which can then be used to set the profile amplitude
as shown in eq.(\ref{eq:finalYscale}). This method is solely dependent on the
cluster size estimate, and is further described in Appendix
\ref{append:selfsim}.

We find agreement between the Coma self-similar SZ model
and the observed \textit{Planck} temperature profiles. Although the
\textit{Planck} data does seem to have both a lower central amplitude and a
slightly wider profile at large angular scales than the self-similar
expectation. We note that corresponding differences between the self-similar
model and the \textit{Planck} data can be seen in the \citet{mozzatta2012}
analysis. A flatter inner profile may also be expected if any pixelisation
effects cause any further smoothing beyond the stated resolution of $10'$. We
also find reasonable agreement between our two estimates of the \textit{Planck}
profile that use different cluster size estimates. Although, as expected, the
model using the \citet{mozzatta2012} value of $R_{500}$ does provide a better
fit to the \textit{Planck} data. We conclude that the agreement between
our \textit{Planck} expectation and the underlying \textit{Planck}
profile supports the validity of our inversion of the \textit{Planck} ESZ
data to obtain \textit{Planck} temperature profiles.

\subsubsection{\citet{pointecouteau2012}}
\label{subsubsec:pointecouteau2012}

\begin{figure}
\begin{center}
  \includegraphics[height=20.0cm]{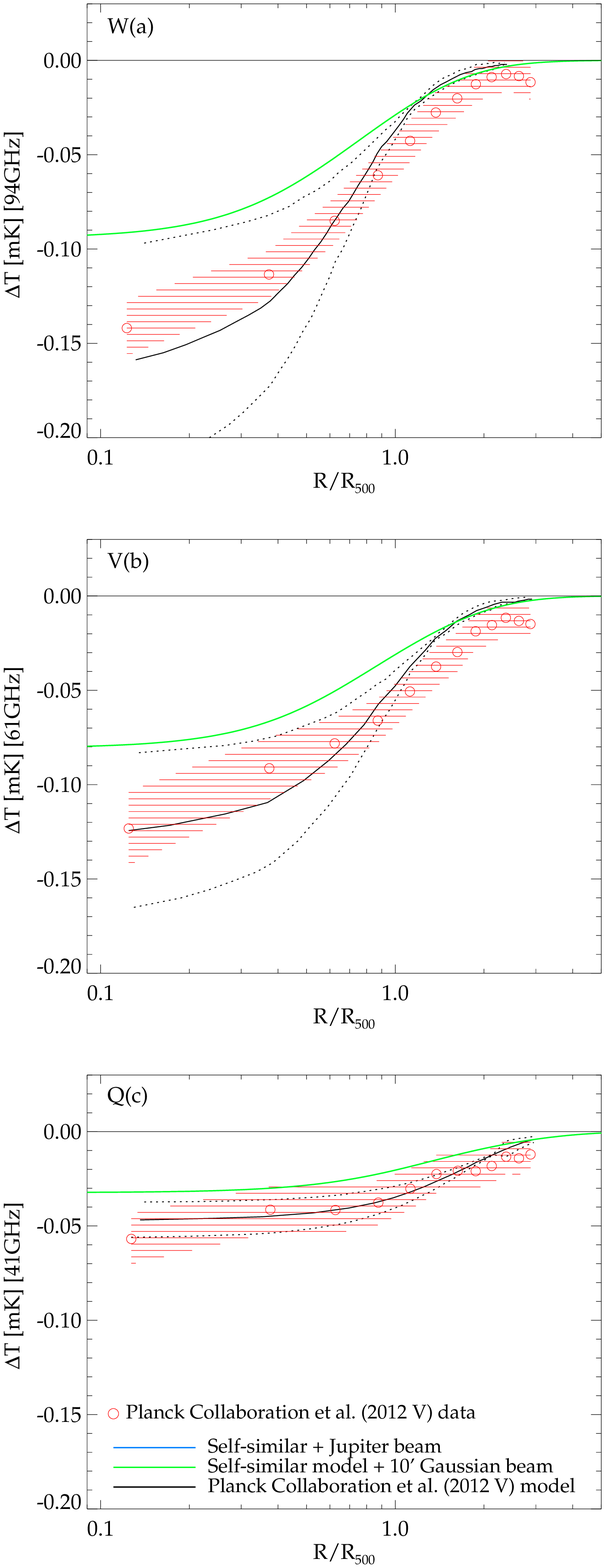}
\caption[]{(a),(b),(c): 
The \textit{WMAP} [W,V,Q] `physical' SZ decrements for the 62
\citet{planckszstudy2011} clusters compared to the \textit{Planck} temperature
decrement from eq.(\ref{eq:selfsimtemppredfinal}). The \textit{Planck} profile
is shown convolved with a $10'$ FWHM Gaussian (green, solid). We also show the
100GHz \textit{Planck} profiles presented in \citet{pointecouteau2012} converted
into thermodynamic temperature at the \textit{WMAP} band centre (red, stripes).
The \citet{pointecouteau2012} models (black, solid) are plotted with their
associated dispersions (black, dotted).
}
 \label{fig:pointecouteau_LSZ62_results}
\end{center}
\end{figure}

In Fig. \ref{fig:pointecouteau_LSZ62_results} we compare our \textit{Planck}
`physical' SZ profiles for the 62 \citet{planckszstudy2011} clusters to the
`physical' \textit{Planck} profiles presented in
\citet{pointecouteau2012}. As was previously shown in \citet{pointecouteau2012}
the \textit{Planck} [100,70,44] GHz profiles are in agreement with the
self-similar expectation. We now expand on this by attempting to use these
results to verify our method of inverting the \textit{Planck} ERCSC data to
obtain \textit{Planck} temperature profiles.

Since we again find that the \citet{planckszstudy2011} estimates of cluster size
can be significantly different from the ESZ estimates we have used an
alternative method of obtaining expected \textit{Planck} temperature profiles.
This method replicates \citet{pointecouteau2012}'s approach in assuming the
\citet{arnaud2010} self-similar model for the cluster and directly evaluating
the Compton-y parameter, as outlined in Appendix \ref{append:selfsim}. We have
further followed \citet{pointecouteau2012}'s Section. 4.3 in using the
\citet{planckszstudy2011} estimates of $\theta_{500}$ and calibrate the central
GNFW pressure, $P_{0}$, using the  X-ray equivalent of the integrated SZ signal,
$Y_{X}$.

As shown in Fig. \ref{fig:pointecouteau_LSZ62_results} the two self-similar
models convolved with $10'$ FWHM Gaussian beam profiles are in agreement beyond
$R_{500}$. However the inner profile of \citet{pointecouteau2012}'s model
(black, solid $\pm$ dotted) is substantially sharper than our model (green,
solid). Although our model lies within the \citet{pointecouteau2012}'s $\approx
1\sigma$ dispersion, we are comparing the stacked models (ie: the statistical
average) so the error range is a $\sqrt{N}\approx8$ smaller. We believe this
difference is caused by the different stacking procedure used in
\citet{pointecouteau2012} where depending on the noise properties within the bin
either logarithmic or linear weightings were used. We have found that using
these alternative weightings can accentuate the central peak of the profile,
although not to the extent necessary for full agreement with
\citet{pointecouteau2012}. We currently do not have a full explanation for the
difference in central amplitude.

\subsection{\textit{WMAP}-\textit{Planck} ESZ comparison}

We next show the stacked \textit{WMAP}7 temperature profiles for 151 
clusters listed in the \textit{Planck} ESZ catalogue. We are using the
`photometric' approach to background subtraction, with an annulus from
$60'-120'$ being used in W (and scaled according to beamwidth in Q
and V). The final models are based on the statistical average of the
clusters.

\begin{figure}
\begin{center}
 
\includegraphics[height=20cm]{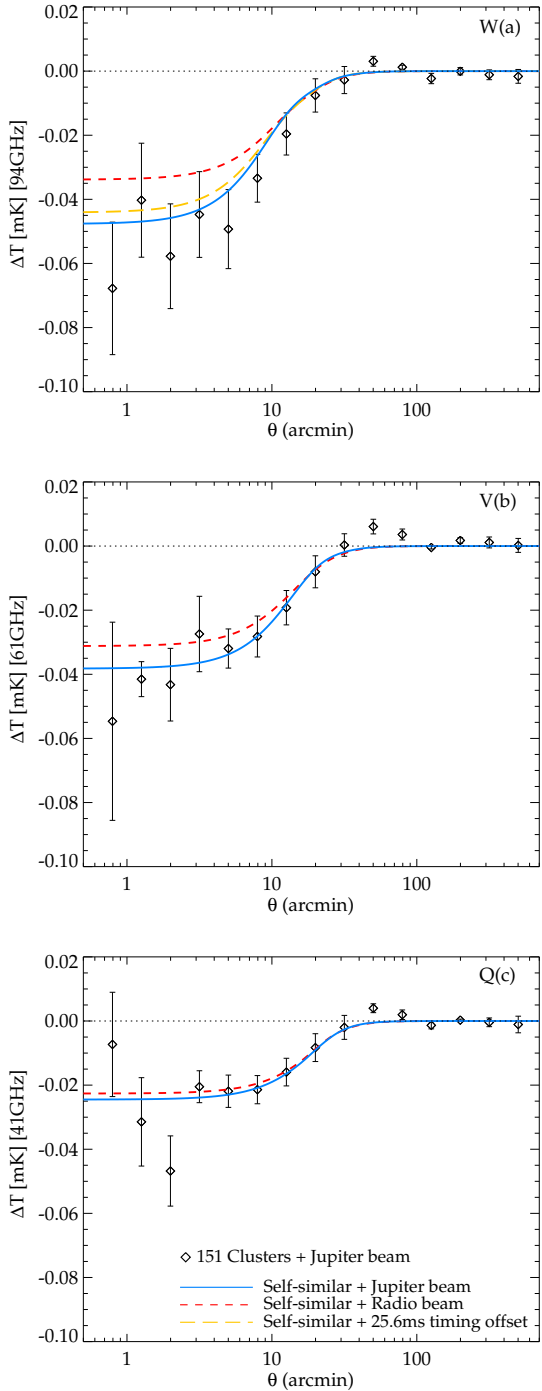}
 \caption[]{(a),(b),(c): 
 The stacked \textit{WMAP} [W,V,Q] SZ decrements for 151 
 \textit{Planck} SZ clusters compared to the stacked \textit{Planck}
temperature decrement from eq.(\ref{eq:selfsimtemppredfinal}). The
\textit{Planck} profile is shown convolved with a \textit{WMAP} Jupiter beam,
a beam fitted to the radio source profiles by \citet{utaneradio} and the
\citet{SawangwitThesis} timing offset derived beam.
 }
\label{fig:sztempcompplanck}
\end{center}
\end{figure}

We see in Fig. \ref{fig:sztempcompplanck} that the \textit{WMAP} data is an
excellent fit to the \textit{Planck} expectation. The fit between the
\textit{Planck} data and the \textit{WMAP} data is not only good in all three
[W,V,Q] bands but at all angular scales. We have further quantified the SZ
measurements using jack-knife and bootstrap techniques all of which support
\textit{WMAP}-\textit{Planck} agreement, however we acknowledge these techniques
don't include covariance.

In Fig. \ref{fig:sztempcompplanck} we have shown the \textit{Planck}
self-similar models convolved with the power-law beams from \citet{utaneradio}.
We find that in the case of the W band where the radio source profiles are most
different from the Jupiter beam, there is now disagreement with the
\textit{WMAP} data with a deficit of $\approx$ 20\% in the centre. In the Q and
V bands where the radio source profiles are closer to the Jupiter
beam, the wider beams give virtually no change in the agreement with the
\textit{WMAP} data. We conclude that the \textit{Planck} SZ profiles disagree
with the \citet{utaneradio} \textit{WMAP}5 radio source profile fits. 

However, the \citet{SawangwitThesis} timing offset derived beam, which provides
an
excellent fit to the radio source profiles as shown in Fig.
\ref{fig:W1_offset_beam}, is significantly less wide than the \citet{utaneradio}
beam. As shown in Fig. \ref{fig:sztempcompplanck} we find that the the timing offset beam only
marginally reduces the central SZ temperature. We therefore conclude that the
\textit{WMAP} SZ results are not at sufficient S/N to differentiate between the
timing offset derived and the Jupiter beams.

\subsection{Coma}
\label{subsec:comasz}

We have also looked at the \textit{Planck} model fits for the Coma
cluster and compared them to \textit{WMAP}. Part of the motivation here is that
previous authors, \citet{lieu2006} and \citet{bielby2007}, have used Coma in
investigating the consistency of the \textit{WMAP} SZ signal with X-ray 
predictions.

\begin{figure}
\begin{center}
  \includegraphics[height=20.0cm]{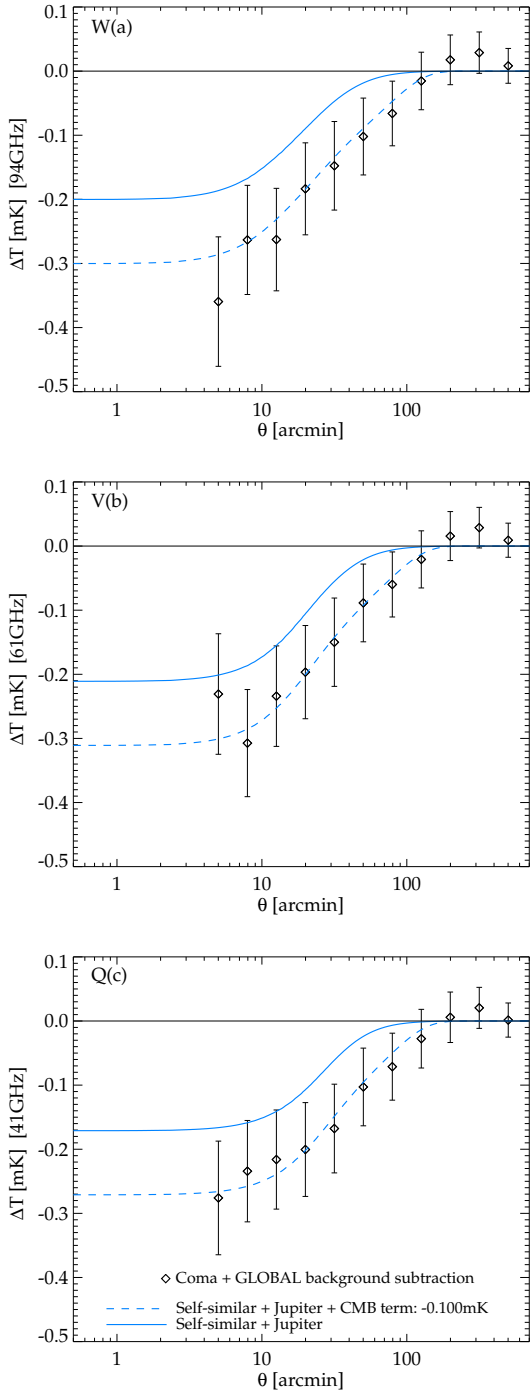}
\caption[]{(a),(b),(c): 
The \textit{WMAP} [W,V,Q] SZ decrements for the Coma cluster compared
to the \textit{Planck} temperature decrement from
eq.(\ref{eq:selfsimtemppredfinal}). The error within each annulus for this
individual cluster is simply the standard deviation of the ESZ clusters and is
therefore only indicative. The \textit{Planck} profile is shown convolved with
a \textit{WMAP} Jupiter beam (blue, solid). We also show a model incorporating a
$100\mu K$ downwards CMB fluctuation (blue, dashed).
}
 \label{fig:mazzotta_coma_results}
\end{center}
\end{figure}

In Fig. \ref{fig:mazzotta_coma_results} we now show the \textit{Planck}
self-similar model for Coma (solid blue line) and see that it is substantially
overestimated by the \textit{WMAP} data. An analogous situation was found by
\citet{wmapkomatsu2011} in that the \textit{WMAP} Coma V and W
band profiles (potentially affected by CMB contamination) showed
$\mathcal{O}(100\mu K)$ more SZ signal than the optimal combined V and W
profiles (free of CMB contamination).

\citet{wmapkomatsu2011} proposed that Coma may sit on $\mathcal{O}(100\mu K)$
downwards CMB fluctuation. Our results are entirely consistent with this
interpretation because the \textit{Planck} MMF method is essentially `CMB-free'
whereas our \textit{WMAP} Coma data may still be contaminated by CMB
fluctuations. On this basis we also show in Fig.
(\ref{fig:mazzotta_coma_results}) a simple alteration to the \textit{Planck}
Coma SZ self-similar model by including a Gaussian with amplitude $-100\mu K$
and $(\mu,\sigma)=(0',60')$ to mimic the effect of a downwards CMB contribution
centred on Coma (blue, dashed). Agreement with the \textit{WMAP} data is
improved when a CMB contamination term is included. We therefore conclude that
the difference between the \textit{Planck} and \textit{WMAP} Coma SZ profiles is
the result of CMB contamination.

\subsection{\citet{bonamente2006} Results}
\label{subsection:bonamente2006res}

\begin{figure}
\begin{center}
  \includegraphics[height=20cm]{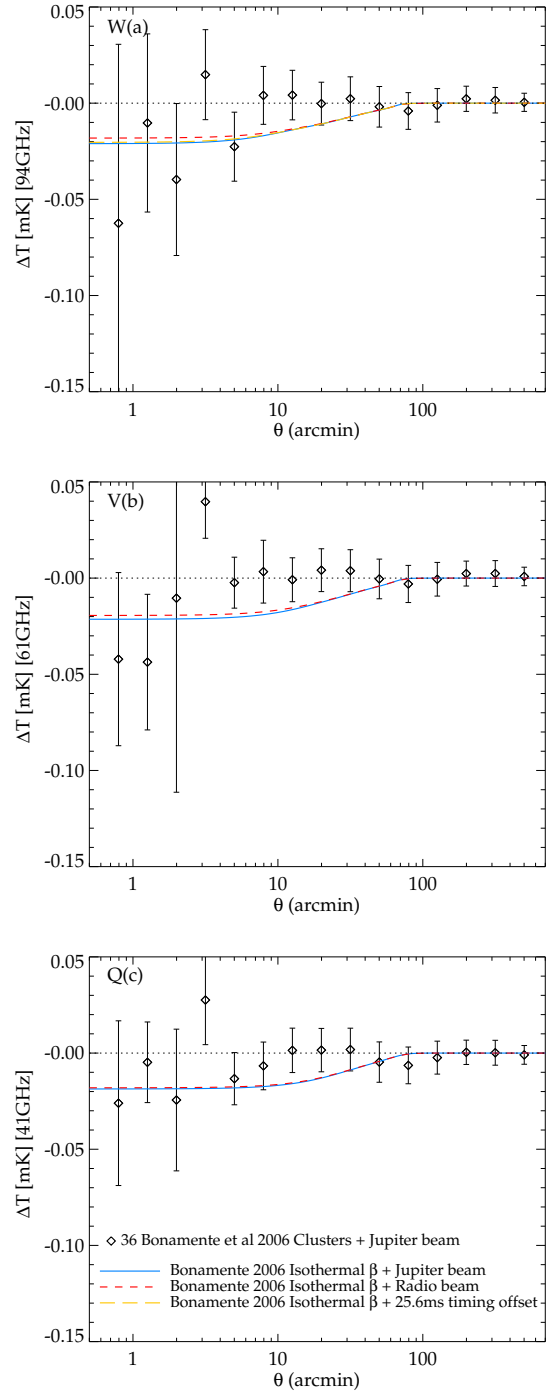}
 \caption[]{(a),(b),(c): 
 The stacked \textit{WMAP} [W,V,Q] SZ decrements for the 36
\citet{bonamente2006} clusters that are unmasked in the \textit{WMAP}
temperature maps. This is compared to a stacked isothermal model based on the
SZ/X-ray parameters fitted by \citet{bonamente2006}, convolved with the Jupiter
beam, a beam fitted to the radio source profiles by \citet{utaneradio} and the
\citet{SawangwitThesis} timing offset derived beam.
 }
\label{fig:scaledsztempcomp_bonamente}
\end{center}
\end{figure}

In \citet{bielby2007} a \textit{WMAP} discrepancy with the SZ/X-ray results of
\citet{bonamente2006} was presented. This is of particular interest as the
\citet{wmapkomatsu2011} \textit{WMAP} SZ discrepancy was largely associated
with the inner profile. The \citet{bonamente2006} results complement this
because they are weighted heavily to small scales because of the high resolution
of their interferometric observations. In Fig.
\ref{fig:scaledsztempcomp_bonamente} we have therefore presented a
stack of the 36 \citet{bonamente2006} clusters that are unmasked in the \textit{WMAP}
temperature maps. We now find good agreement between
the \textit{WMAP} SZ decrements and the \citet{bonamente2006} SZ/X-ray expectation.
This is in contradiction to the results of \citet{bielby2007}. We have found
this is attributable to \citet{bielby2007}'s assumption that the
cluster is well resolved with respect to the \textit{WMAP} beam. As discussed in
Section \ref{subsec:model_sz_convol} this assumption introduces a systematic
error into their 1-D convolution with the \textit{WMAP} beam profiles. We
therefore now report no evidence for a \textit{WMAP} SZ discrepancy with respect
to the \citet{bonamente2006} X-ray models.

\section{Discussion}
\label{sec:discussion}

The main criticism that was made of the previous results of
\citet{utaneradio} was that the wide \textit{WMAP} radio source profiles may be
caused by Eddington bias \citep{eddington1913}. Essentially, low S/N
sources detected in the \textit{WMAP} data may be contaminated by upwards CMB
fluctuations and not balanced by downwards fluctuations. This could
explain the wider than expected profiles, particularly at faint fluxes.

There may be some evidence for Eddington bias in the faintest \textit{WMAP}5 W
band source sub-sample that was initially used by \citet{utaneradio}. However,
it was because of this the faintest sources were not used in \citet{utaneradio}
fits of the beam profile and that a flux limit of $S\geq1.1$Jy has been used in
calculating our radio source profiles. We also note that the \textit{Planck}
sources show the wider beam independent of whether the CMBSUBTRACT flag applies.
We further note that we have restricted the \textit{Planck} sources to have a
FWHM strictly less than the \textit{WMAP} W band beam profile width and
find a wider than expected beam profile for these clearly point
sources. \textit{WMAP} sources selected from a `CMB-free' map and NVSS selected
sources at low frequency also show the same wider than expected beam.

Furthermore, we have also run  Monte Carlo re-simulations of the source
detection, producing artifical source catalogues extracted from
simulated CMB maps. Here, after applying the same cross-correlation
technique as for the data, the \textit{WMAP} beam was recovered as input
(see Fig. \ref{fig:sim}), again arguing that these sources are little
affected by Eddington bias.

The \textit{Planck} data also confirms the non-linearity of
\textit{WMAP} fluxes, particularly in the W band, in the range previously used.
Decreasingly non-linear effects are also seen at Q and V. But when ground-based
and \textit{Planck} data for the bright \citet{weiland2011} sources are included
in these comparisons the evidence for non-linearity becomes less and it could
still be that the discrepancy corresponds more to a constant offset.

Given that the beam profile widening is smaller in the \textit{WMAP}7 data than
in \textit{WMAP}5, a scan pattern timing offset as discussed by \citet{liuli}
becomes a more plausible explanation for this effect.  We have seen that the
effect, originally invoked as an explanation for the alignment of the low order
multipoles with the ecliptic, also provides a reasonable fit to the W band beam
profiles (see Fig. \ref{fig:W1_offset_beam}).

In our comparison of \textit{Planck}-\textit{WMAP} SZ decrements we have found
good agreement. Similarly, our \textit{WMAP} SZ profile results
are now in agreement with the X-ray data for the \citet{bonamente2006} sample. This
work is now in line with previous authors who when studying the integrated
\textit{WMAP} SZ signal $Y_{tot}$ (as opposed to the Compton y-parameter) have found no evidence for a \textit{WMAP} discrepancy
\citep{melin2011}. We have no explanation for the \citet{wmapkomatsu2011}
\textit{WMAP} SZ profile discrepancies at this point.

We have also found our \textit{Planck} profiles are consistent with the
\textit{Planck} results of \citet{pointecouteau2012} and \citet{mozzatta2012}.
We interpret this as validating our our method of inverting the \textit{Planck}
ERCSC data to obtain \textit{Planck} SZ temperature profiles.

In response to a question from a referee, we note the Integrated Sachs-Wolfe 
(ISW) effect is at most a 10$\mu K$ effect for clusters/superclusters,
\citep{granett2008}. This is too marginal to affect the profiles we recover. The
ISW is an even more negligible effect for radio sources, as observed by
\citep{nolta2004,sawangwit2010_ISW} where it is shown to be
$\approx$ 0.3$\mu K$ effect. It is therefore highly unlikely to cause any 
bias to our results in either the SZ or radio source analyses.

\begin{figure*}
 \includegraphics[width=17cm]{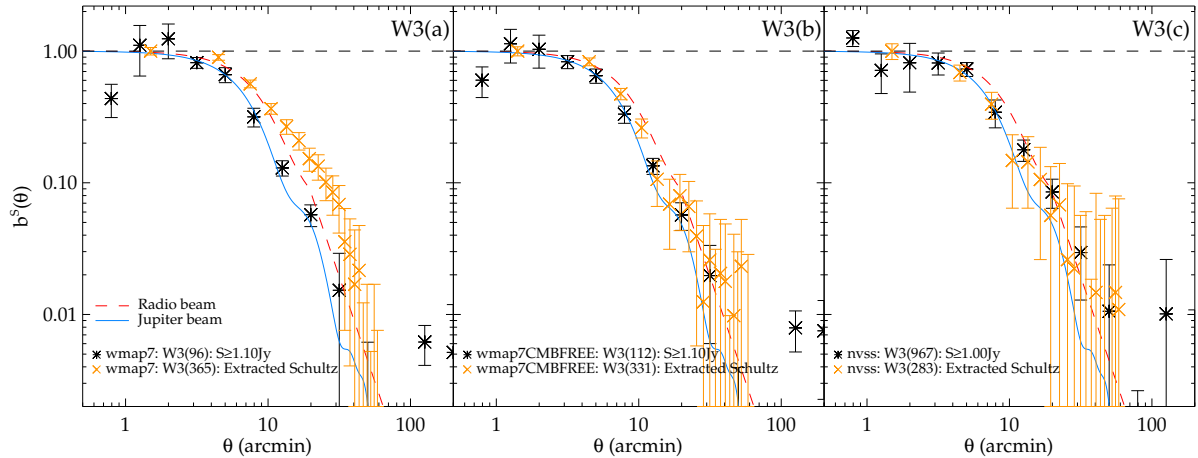}
 \caption[]{(a),(b),(c): The photometrically subtracted stacked
 \textit{WMAP}7 W3 $b^{s}(\theta)$ profiles for \textit{WMAP}7, \textit{WMAP}7
CMB-free and NVSS catalogues as compared to the corresponding W3
 results from \citet{schultz2011} as taken from their Fig. 5. Also shown
 are the $b^{s}(\theta)$ for the Jupiter beam (blue, solid) and the
 radio source fit (red, dashed) of \citet{utaneradio}.
 }
 \label{fig:radio_schultzcompare}
\end{figure*}

We have also compared our results to those of \citet{schultz2011} whose paper
appeared while this one was being refereed. We compare our results directly to
theirs in Fig. \ref{fig:radio_schultzcompare}. The \textit{WMAP}7 W3 graph they
use as an example is significantly wider than any profile shown by
\citet{utaneradio} or here. This is because they have used a sample with no cut
at all in terms of significance of detection or flux  and clearly these data
will be strongly affected by Eddington bias. We repeat that at the flux limits
used here, the simulations show no sign of such bias and so we are confident
that this criticism does not apply to our results.  We note that there are
additional quality cuts that \citet{schultz2011} have made with respect to our
work, such as an expanded mask and a culling of close pairs. However, we find that our results are unchanged when we apply them as well. We find that
their \textit{WMAP}7-CMBfree and NVSS beam profiles are very comparable to ours
for the W3 band and they are wider than the Jupiter profile as can be seen.
\citet{schultz2011} suggest that the \textit{WMAP}7-CMBfree profiles are wider
due to errors on the radio source positions. However, their assumed positional
errors may be overestimates for their stacked radio source profiles since the
stacks are weighted towards the brighter radio sources which have more accurate
positions. The fact that we are using 5GHz GB6 and PMN positions accurate to
sub-$0.'5$ accuracy in the \textit{WMAP} case and obtain
\textit{WMAP}7-CMBfree profiles consistent with \citet{schultz2011} suggests that
positional errors cannot be the explanation. The main difference with the NVSS
results of \citet{schultz2011} is their larger errors. Our NVSS sample is
$\approx4$x larger than theirs due to our 1.4GHz flux limit of 1Jy compared to
their 2Jy limit, this (as well as our larger binning) explains most of the
difference in errors. Otherwise the results appear entirely consistent.

\begin{figure}
\begin{center}
   \includegraphics[width=7.5cm]{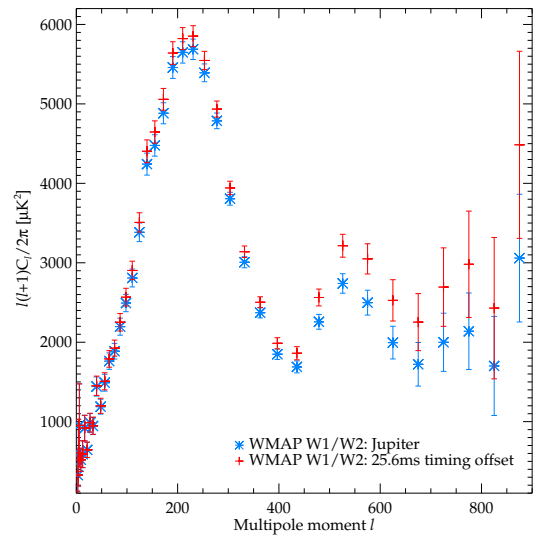}
 \caption[]{The \textit{WMAP} W1/W2 $C_l$ result from debeaming with the Jupiter
beam (blue, stars) as compared to the result from debeaming with the
timing offset derived beam from \citet{SawangwitThesis} and also shown in Fig.
\ref{fig:W1_offset_beam} (red, crosses).
}
 \label{fig:powerspectrafluxcorrectW}
\end{center}
\label{fig:nl_cl}
\end{figure}

We finally show in Fig. \ref{fig:powerspectrafluxcorrectW} the effect
wide \textit{Planck} radio source profiles (see Fig.
\ref{fig:radioproftempandbsplanck}f) has on the \textit{WMAP} W1/W2 $C_l$. We
take here the model with the 25.6ms timing offset that gave a reasonable fit to
the data in Fig. \ref{fig:W1_offset_beam}. Otherwise without the model, we would
need to extrapolate a fit out to large angles. Then debeaming the raw W1/W2
$C_l$ from PolSpice \citep{szapudi2001} via eq.(1,2) of \citet{utaneradio}, we
see that the $C_{\ell}$ shows a modest increase in amplitude at $\ell<$400, with
a larger increase at $\ell>$400. This reflects where the Jupiter and timing
offset beam are different. At $\ell<$400 there is very little difference between
the Jupiter and the timing offset beam. We note that this region is essentially
unconstrained by the radio source profiles. So the lack of change to the first
acoustic peak is tied to the specific details of the timing offset model. A
different model could give significantly different results and therefore
\textit{WMAP}7 first acoustic peak's amplitude and position relies heavily on
the accuracy of the observed Jupiter beam beyond 30$'$ scales, which is untested
by our observations.


\section{Conclusions}
\label{sec:conclusions}

We have investigated the beam profile of \textit{WMAP} by comparing
beam profiles from radio sources with the Jupiter beam profile. We have
compared sources from \textit{Planck}, \textit{WMAP}, \textit{WMAP} CMB-free
and NVSS catalogues. We find that in all cases the radio sources show wider
profiles than the Jupiter beam with little indication of Eddington bias
or dependence on the method of normalisation. Applying our
cross-correlation to realistic simulations strongly supports the
accuracy of our beam profile measurements. However, it must be said that
in the \textit{WMAP}7 data the W radio source profiles are less wide
than previously found  by \citet{utaneradio} in the \textit{WMAP}5
release. The rejection of the Jupiter beam is now only $\approx3\sigma$
in the \textit{Planck} radio source comparison. But the rejection of the
Jupiter beam is reasonably consistent between the admittedly overlapping
radio source samples from \textit{Planck}, \textit{WMAP} and NVSS. We
have therefore considered explanations for the wide profiles assuming 
that they are not statistical fluctuations. Two such possibilities are a
non-linearity in the \textit{WMAP} temperature scale and a timing offset in the
\textit{WMAP} scan pattern as discussed by \citet{liuli}. The narrower profiles
measured here compared to the \citet{utaneradio} \textit{WMAP}5 profiles
increase the possibility of their being explained by a timing offset.

We have also found discrepancies between \textit{WMAP} fluxes compared to
\textit{Planck} and ground-based fluxes. For $S<30$Jy the \textit{WMAP} fluxes
look to have a non-linear relation with \textit{Planck} fluxes. However, when
the further very bright sources discussed by \citet{weiland2011} with
ground-based and \textit{Planck} measurements are included then this flux-flux
discrepancy appears more like a linear than a non-linear offset.

We have compared stacked \textit{WMAP} SZ decrements with those
measured by \textit{Planck} and by ground-based observations. In contrast to
previous reports we now find \textit{WMAP} agrees with both the
\textit{Planck} and ground-based data. However this work is not at high enough
S/N to distinquish between the timing offset beam of \citet{SawangwitThesis} and
the \textit{WMAP} Jupiter beam.

We have shown that transforming the Jupiter beam using a model that fits
the radio source profiles results in small but significant changes to
the \textit{WMAP} $C_{\ell}$. At the least, a wider beam would imply a
much larger uncertainty in the normalisation and hence the estimate of
$\sigma_8$ from \textit{WMAP}. Unfortunately, faint radio sources cannot
check the \textit{WMAP} beam at scales larger than $30'$ and a wider beam at
these scales could, in principle, change the position, as well as the
normalisation, of even the first acoustic peak. Clearly it is important
to continue to test the calibration and beam profile of \textit{WMAP},
particularly in the W band.


\section*{Acknowledgments}


JRW acknowledges financial support from STFC. US acknowledges financial
support from the Royal Thai Government.
We acknowledge the use of data from NASA \textit{WMAP} and ESA \textit{Planck}
collaborations.






\setlength{\bibhang}{2.0em}
\setlength\labelwidth{0.0em}

\bibliographystyle{mn2e}

\begin{thebibliography}{}


\bibitem[\protect\citeauthoryear{Arnaud et 
al.}{2010}]{arnaud2010} Arnaud M., Pratt G.~W., Piffaretti R., B{\"o}hringer H., Croston J.~H., Pointecouteau E., 2010, A\&A, 517, A92 

\bibitem[Bailey 
\& Sparks(1983)]{baileysparks1983} Bailey, M.~E., \& Sparks, W.~B.\ 1983, \mnras, 204, 53P 

\bibitem[Bennett et al.(2003)]{bennett2003} Bennett, C.~L., et al.\ 
2003, \apjs, 148, 97 


\bibitem[\protect\citeauthoryear{Bennett et 
al.}{2011}]{bennett2011} Bennett C.~L., et al., 2011, ApJS, 192, 17 


\bibitem[\protect\citeauthoryear{Bielby 
\& Shanks}{2007}]{bielby2007} Bielby R.~M., Shanks T., 2007, MNRAS, 382, 1196


\bibitem[\protect\citeauthoryear{Blake 
\& Wall}{2002}]{blake2002} Blake C., Wall J., 2002, MNRAS, 337, 993 


\bibitem[\protect\citeauthoryear{Bonamente et 
al.}{2006}]{bonamente2006} Bonamente M., Joy M.~K., LaRoque S.~J., 
Carlstrom J.~E., Reese E.~D., Dawson K.~S., 2006, ApJ, 647, 25 


\bibitem[\protect\citeauthoryear{Cavaliere 
\& Fusco-Femiano}{1976}]{cavaliere1976} Cavaliere A., Fusco-Femiano R., 1976, A\&A, 49, 137 

\bibitem[\protect\citeauthoryear{Chen 
\& Wright}{2009}]{chen2009} Chen X., Wright E.~L., 2009, ApJ, 694, 222 

\bibitem[Condon et al.(1998)]{Condon98} Condon, J.~J., Cotton, 
W.~D., Greisen, E.~W., et al.\ 1998, \aj, 115, 1693 

\bibitem[Cotton et al.(2009)]{cotton2009} Cotton, W.~D., Mason, 
B.~S., Dicker, S.~R., et al.\ 2009, \apj, 701, 1872

\bibitem[\protect\citeauthoryear{Eddington}{1913}]{eddington1913} 
Eddington A.~S., 1913, MNRAS, 73, 359 

\bibitem[\protect\citeauthoryear{Gold et al.}{2011}]{gold2011} 
Gold B., et al., 2011, ApJS, 192, 15 

\bibitem[\protect\citeauthoryear{Granett, Neyrinck, 
\& Szapudi}{2008}]{granett2008} Granett B.~R., Neyrinck M.~C., Szapudi I., 2008, ApJ, 683, L99 

\bibitem[\protect\citeauthoryear{Gregory et 
al.}{1996}]{gregory1996} Gregory P.~C., Scott W.~K., Douglas K., 
Condon J.~J., 1996, ApJS, 103, 427 

\bibitem[\protect\citeauthoryear{Griffith 
\& Wright}{1993}]{griffith1993} Griffith M.~R., Wright A.~E., 1993, AJ, 105, 1666 

\bibitem[\protect\citeauthoryear{Hill et al.}{2009}]{hill2009} 
Hill R.~S., et al., 2009, ApJS, 180, 246 

\bibitem[Hinshaw et al.(2003)]{hinshaw03} Hinshaw, G., Spergel, 
D.~N., Verde, L., et al.\ 2003, \apjs, 148, 135 

\bibitem[\protect\citeauthoryear{Jarosik et 
al.}{2003}]{jarosik2003} Jarosik N., et al., 2003, ApJS, 145, 413 

\bibitem[\protect\citeauthoryear{Jarosik et 
al.}{2011}]{jarosik2011} Jarosik N., et al., 2011, ApJS, 192, 14 

\bibitem[Kenney 
\& Dent(1985)]{kenney1985} Kenney, J.~D., \& Dent, W.~A.\ 1985, \apj, 298, 644 

\bibitem[\protect\citeauthoryear{Komatsu et 
al.}{2011}]{wmapkomatsu2011} Komatsu E., et al., 2011, ApJS, 192, 18 

\bibitem[\protect\citeauthoryear{Lieu, Mittaz, 
\& Zhang}{2006}]{lieu2006} Lieu R., Mittaz J.~P.~D., Zhang S.-N., 2006, ApJ, 648, 176 


\bibitem[\protect\citeauthoryear{Limon et al.}{2008}]{Limon08}
Limon M., et al., 2008, Wilkinson Microwave Anisotropy Probe (\textit{WMAP}): Five Year Explanatory Supplement, \texttt{http://lambda.gsfc.nasa.gov/data/map/doc/}
\texttt{MAP\_supplement.pdf}


\bibitem[\protect\citeauthoryear{Liu 
\& Li}{2011}]{liuli} Liu H., Li T.-P., 2011, ApJ, 732, 125 

\bibitem[Liszt 
\& Lucas(1999)]{liszt1999} Liszt, H., \& Lucas, R.\ 1999, \aap, 347, 258 

\bibitem[Lonsdale et al.(1998)]{lonsdale1998} Lonsdale, C.~J., 
Doeleman, S.~S., \& Phillips, R.~B.\ 1998, \aj, 116, 8 


\bibitem[\protect\citeauthoryear{Malik et al.}{in prep}]{malik11}
Malik, Sawangwit U., Shanks T. \& Whitbourn, J.R., 2011, In preparation

\bibitem[\protect\citeauthoryear{Melin et 
al.}{2011}]{melin2011} Melin J.-B., Bartlett J.~G., Delabrouille J., Arnaud M., Piffaretti R., Pratt G.~W., 2011, A\&A, 525, A139

\bibitem[\protect\citeauthoryear{Melin, Bartlett, 
\& Delabrouille}{2006}]{melin2006} Melin J.-B., Bartlett J.~G., Delabrouille J., 2006, A\&A, 459, 341 

\bibitem[\protect\citeauthoryear{Moss, Scott, 
\& Sigurdson}{2011}]{moss2011} Moss A., Scott D., Sigurdson K., 2011, JCAP, 1, 1 

\bibitem[\protect\citeauthoryear{Mroczkowski et 
al.}{2009}]{southpole2009} Mroczkowski T., et al., 2009, ApJ, 694, 
1034 

\bibitem[\protect\citeauthoryear{Myers et al.}{2004}]{myers2004} 
Myers A.~D., Shanks T., Outram P.~J., Frith W.~J., Wolfendale A.~W., 2004, 
MNRAS, 347, L67 

\bibitem[\protect\citeauthoryear{Nagai, Vikhlinin, 
\& Kravtsov}{2007}]{nagai2007} Nagai D., Vikhlinin A., Kravtsov A.~V., 2007, ApJ, 655, 98 

\bibitem[\protect\citeauthoryear{Nolta et al.}{2004}]{nolta2004} 
Nolta M.~R., et al., 2004, ApJ, 608, 10 

\bibitem[\protect\citeauthoryear{Page et al.}{2003}]{page2003} 
Page L., et al., 2003, ApJS, 148, 39 

\bibitem[\protect\citeauthoryear{Page et al.}{2003}]{page2003preflight} 
Page L., et al., 2003, ApJ, 585, 566 

\bibitem[\protect\citeauthoryear{Piffaretti et 
al.}{2005}]{piffaretti2005} Piffaretti R., Jetzer P., Kaastra J.~S., Tamura T., 2005, A\&A, 433, 101 

\bibitem[\protect\citeauthoryear{\textit{Planck} Collaboration et 
al.}{2011d}]{planckearlySZ} \textit{Planck} Collaboration et al, 2011d, The all-sky early Sunyaev-Zeldovich cluster sample, arXiv, 
arXiv:1101.2024 

\bibitem[\protect\citeauthoryear{\textit{Planck} Collaboration et 
al.}{2011f}]{planckxraystats2011} \textit{Planck} Collaboration et al, 2011f, Statistical analysis of Sunyaev-Zeldovich scaling relations for X-ray galaxy clusters, arXiv, arXiv:1101.2043 

\bibitem[\protect\citeauthoryear{\textit{Planck} Collaboration et 
al.}{2011g}]{planckszstudy2011} \textit{Planck} Collaboration et al, 2011g, Calibration of the local galaxy cluster Sunyaev-Zeldovich scaling relations, arXiv, arXiv:1101.2026

\bibitem[\protect\citeauthoryear{\textit{Planck} Collaboration et
al.}{2011exp}]{planckERCSCexplain} \textit{Planck} Collaboration et al, 2011exp, Explanatory supplement,
\texttt{http://www.sciops.esa.int/SA/PLANCK/docs/}
\texttt{ERCSC\_Explanatory\_Supplement.zip}

\bibitem[\protect\citeauthoryear{\textit{Planck} Collaboration et 
al.}{2012V}]{pointecouteau2012} \textit{Planck} Collaboration et al, 2012V, Pressure profiles of galaxy clusters from the Sunyaev-Zeldovich effect, arXiv, arXiv:1207.4061 

\bibitem[\protect\citeauthoryear{\textit{Planck} Collaboration et 
al}{2012X}]{mozzatta2012} \textit{Planck} Collaboration et al, 2012X, Physics of the hot gas in the Coma cluster, arXiv, arXiv:1208.3611 

\bibitem[\protect\citeauthoryear{Pratt et 
al.}{2007}]{pratt2007} Pratt G.~W., B{\"o}hringer H., Croston J.~H., Arnaud M., Borgani S., Finoguenov A., Temple R.~F., 2007, A\&A, 461, 71 

\bibitem[\protect\citeauthoryear{Press et al.}{1992}]{press1993} 
Press W.~H., Teukolsky S.~A., Vetterling W.~T., Flannery B.~P., 1992, In `Numerical Recipes', CUP:Cambridge, pp. 660

\bibitem[\protect\citeauthoryear{Refregier, Spergel, 
\& Herbig}{2000}]{refregier2000} Refregier A., Spergel D.~N., Herbig T., 2000, ApJ, 531, 31 

\bibitem[\protect\citeauthoryear{Roukema}{2010}]{roukema2011} 
Roukema B.~F., 2010, arXiv, arXiv:1007.5307 

\bibitem[Salter et al.(1989)]{salter1989} Salter, C.~J., Reynolds, 
S.~P., Hogg, D.~E., Payne, J.~M., \& Rhodes, P.~J.\ 1989, \apj, 338, 171 

\bibitem[\protect\citeauthoryear{Sawangwit \& Shanks}{2010a}]{utaneradio}
Sawangwit U., Shanks T., 2010a, MNRAS, 407, L16

\bibitem[\protect\citeauthoryear{Sawangwit \& Shanks}{2010b}]{tommoriond}
Sawangwit U., Shanks T., 2010b, In `45th Rencontres de Moriond: Cosmology
2010', Eds Auge, E., Dumarchez, J. \& Tran Thanh Van, J., pp. 53-57, GIOI:
Vietnam, (arXiv:1006.1270).

\bibitem[\protect\citeauthoryear{Sawangwit et 
al.}{2010}]{sawangwit2010_ISW} Sawangwit U., Shanks T., Cannon R.~D., 
Croom S.~M., Ross N.~P., Wake D.~A., 2010, MNRAS, 402, 2228 


\bibitem[\protect\citeauthoryear{Sawangwit}{2011}]{SawangwitThesis} Sawangwit U., 2011, PhD thesis.

\bibitem[Schultz 
\& Huffenberger(2011)]{schultz2011} Schultz, K.~W., \& Huffenberger, K.~M.\ 2011, arXiv:1111.7311 

\bibitem[\protect\citeauthoryear{Shanks}{1985}]{shanks85} Shanks 
T., 1985, Vistas in Astr., 28, 595

\bibitem[\protect\citeauthoryear{Sunyaev 
\& Zeldovich}{1980}]{sunyaev1980} Sunyaev R.~A., Zeldovich I.~B., 1980, ARA\&A, 18, 537 

\bibitem[\protect\citeauthoryear{Szapudi, Prunet, 
\& Colombi}{2001}]{szapudi2001} Szapudi I., Prunet S., Colombi S., 2001, ApJ, 561, L11 

\bibitem[\protect\citeauthoryear{Tegmark 
\& de Oliveira-Costa}{1998}]{Tegmark98} Tegmark M., de Oliveira-Costa A., 1998, \apjl, 500, L83

\bibitem[\protect\citeauthoryear{Weiland et 
al.}{2011}]{weiland2011} Weiland J.~L., et al., 2011, ApJS, 192, 19 

\bibitem[\protect\citeauthoryear{Wright et al.}{2009}]{wright2009} 
Wright E.~L., et al., 2009, ApJS, 180, 283 

\bibitem[Wright 
\& Birkinshaw(1984)]{wright_and_birkinshaw1984} Wright, M., \& Birkinshaw, M.\ 1984, \apj, 281, 135 

\bibitem[Wright 
\& Sault(1993)]{wright_sault1993} Wright, M.~C.~H., \& Sault, R.~J.\ 1993, \apj, 402, 546 


\end{thebibliography}


\appendix

\section{SZ Self Similar Model}
\label{append:selfsim}

In the self-similar SZ model as employed in  the \textit{Planck} ESZ,  
the fundamental parameters of a cluster are $P_{500}$, $M_{500}$ 
and $R_{500}$. Using  the terminology of \citet{arnaud2010},

\begin{align}
 M_{500} &= \frac{4\pi}{3} R_{500}^{3} 500 \rho_{crit} \label{eq:mass500groundup}, \\
 R_{500} &= D_{a}(z) \frac{\theta_{5R_{500}}}{5}. \label{eq:theta500groundup}
\end{align}

\noindent A $Y_{500}$ parameter corresponding to these is also defined,

\begin{equation}
 Y_{500} = \frac{\sigma_{t}}{m_{e}c^{2}} \frac{4\pi R^{3}_{500}}{3} P_{500},
\label{eq:Y500def}
\end{equation}

\noindent which can be used as a characteristic SZ parameter instead of
$P_{500}$. In eq.({\ref{eq:Y500def}}) the units of $Y_{500}$ are Mpc$^2$, but
are easily convertible to the arcmin$^2$ units used in the ESZ and throughout this
paper\footnote{$Y[Mpc^{2}] = \frac{1}{60^{2}}
(\frac{\pi}{180})^{2}(D_{a}[Mpc])^{2} Y[arcmin^{2}]$.}. This $Y_{500}$ is
a distinct quantity from $Y(R_{500})$ as found by evaluating eq.
({\ref{eq:rawcylprofile}}). The introduction of $Y_{500}$ is well motivated
because, as shown by \citet{arnaud2010}, it allows a scale-free description of
eq.({\ref{eq:rawcylprofile}})'s $Y_{sph}$ and $Y_{cyl}$ in terms of
$x=R/R_{500}$ as follows,

\begin{align}
 Y_{sph}(x) &= Y_{500}  I(x),\\
 Y_{cyl}(x) &= Y_{sph}(5R_{500}) - Y_{500} J(x).
\label{eq:cylprofile}
\end{align}

\noindent where $I(x)$ and $J(x)$ are the spherical and cylindrical scaling functions,

\begin{align}
 I(x) &= \int^{x}_{0} 3 \mathcal{P}(u) u^{2} du, \\
 J(x) &= \int^{5}_{x} 3 \mathcal{P}(u) (u^{2}-x^{2})^{1/2} u du.
\label{eq:jscaling}
\end{align}

\noindent We therefore find that
\begin{equation}
 Y_{cyl}(x) = Y_{500} (I(5)-J(x)).
\label{eq:finalYscale}
\end{equation}

Finally, we can use the above to calculate $Y_{cyl}(R)$ and the Compton
\textit{y} parameter, where $y(\theta) = \frac{d}{d\Omega}Y_{cyl}(\theta)$. We
now describe three methods for doing so.\\

\noindent \textbf{1. Using $Y(5R_{500})$ as an amplitude}: Since $Y_{cyl}(5) =
Y_{sph}(5) = I(5)Y_{500}$, eq.(\ref{eq:finalYscale}) can be expressed as,
\begin{equation}
 Y_{cyl}(R) = Y_{cyl}(5R_{500}) \bigg( 1 - \frac{J(x)}{I(5)} \bigg).
\label{eq:thereallyusefulY2}
\end{equation}
This is the method we adopt in this paper, it is dependent on both $Y(5R_{500})$
and $\theta_{5R500}$.

\noindent \textbf{2. Using $Y_{500}$ as an amplitude}: $Y_{500}$ can be
calculated using $M_{500}$ and $P_{500}$. We can therefore directly evaluate
$Y_{cyl}(R)$ using eq.(\ref{eq:finalYscale}). This method is independent of the
\textit{Planck} provided $Y(5R_{500})$ and instead solely uses $\theta_{5R500}$.

\noindent \textbf{3. Avoiding the integrated SZ signal}: The Compton y-parameter
can be expressed as \citep{pointecouteau2012}, 
\begin{equation}
 y(r) = \frac{\sigma_{t}}{m_{e}c^{2}} \int^{R_{tot}}_{r} \frac{2P(r')r'dr'}{(r'^{2}-r^{2})^{1/2}}.
 \label{eq:ytheta_cylprofile}
\end{equation}
We can therefore directly evaluate the Compton y-parameter if a self-similar
cluster profile is assumed for $P(r)$. This method is independent of the
\textit{Planck} provided $Y(5R_{500})$ and instead solely uses $\theta_{5R500}$.


\bsp

\label{lastpage}


\end{document}